\documentclass[11pt]{article}
\pdfoutput=1

\usepackage{amsmath}
\usepackage{amssymb}
\usepackage{amsfonts}
\usepackage{mathrsfs}

\usepackage{mathrsfs}
\usepackage{fullpage}
\usepackage{setspace}
\usepackage{bm}
\usepackage{bbm}
\usepackage{relsize}
\usepackage{adjustbox}
\usepackage{multirow}
\usepackage{cite}
\usepackage{filecontents}

\usepackage[normalem]{ulem}
\usepackage{enumerate}
\usepackage{yfonts}

\usepackage{psfrag}
\usepackage{array}
\usepackage{booktabs}
\newcolumntype{N}{>{\centering\arraybackslash}m{.5in}}
\newcolumntype{G}{>{\centering\arraybackslash}m{2in}}

\usepackage[symbol]{footmisc}

\usepackage[latin1]{inputenc}
\usepackage{graphicx}
\usepackage{cancel}
\usepackage{slashed}
\usepackage{mathtools}

\usepackage[font=small,labelfont=bf]{caption}
\usepackage{subcaption}

\usepackage[colorlinks=true]{hyperref} 
\hypersetup{
    unicode=false,          
    pdftoolbar=true,        
    pdfmenubar=true,        
    pdffitwindow=false,     
    pdfstartview={FitH},    
    pdftitle={Eccentric extreme mass-ratio inspirals: A gateway to probe quantum gravity effects},    
    pdfauthor={Tieguang Zi, Shailesh Kumar},     
    pdfnewwindow=true,      
    colorlinks=true,       
    linkcolor=blue,          
    citecolor=red,        
    filecolor=magenta,      
    urlcolor=blue           
    }

\usepackage{hyperref}
\hypersetup{colorlinks=true}

\def\equationautorefname~#1\null{%
	Eq.~(#1)\null
}
\def\figureautorefname~#1\null{%
	Fig.~#1\null
}
\def\tableautorefname~#1\null{%
	Table.~#1\null
}
\def\sectionautorefname~#1\null{%
	Section #1\null
}
\def\appendixautorefname~#1\null{%
	Appendix #1\null
}

\begin{document}

\numberwithin{equation}{section}
{
\begin{titlepage}
\begin{center}

\hfill \\
\hfill \\
\vskip 0.75in

{\Large {\bf Eccentric extreme mass-ratio inspirals: A gateway to probe quantum gravity effects
}}

\vskip 0.24in

{\large Tieguang Zi${}$\footnote[2]{\href{mailto:zitg@scut.edu.cn}{zitg@scut.edu.cn}}$^{a,b}$, Shailesh Kumar${}$\footnote[1]{\href{mailto: shailesh.k@iitgn.ac.i}{shailesh.k@iitgn.ac.in}}$^{c}$}
\vspace{0.42cm}

{\it ${}$$^a$ Department of physics, Nanchang University,
	Nanchang 330031, China}

{\it ${}$$^b$ Center for Relativistic Astrophysics and High Energy Physics, Nanchang University, Nanchang, 330031, China}

\vskip 0.05in

{\it ${}$$^c$Indian Institute of Technology, Gandhinagar, Gujarat-382355, India}

\vskip.5mm

\end{center}

\vskip 0.35in

\begin{center} 
{\bf ABSTRACT }
\end{center}
We examine a loop quantum gravity (LQG) inspired rotating black hole, treating it as a central supermassive black hole (SMBH) in an extreme mass-ratio inspiral (EMRI) system, where an inspiralling object exhibits eccentric motion around the SMBH. With the orbital dynamics, we derive analytical expressions for the rate of change of orbital energy and angular momentum, as well as orbital evolution, and subsequently generate the gravitational waveforms. To evaluate the difference between EMRI waveforms emitted from the Kerr black hole and a spinning black hole in LQG, we compute the dephasing and mismatch using the Laser Interferometer Space Antenna (LISA) observation. Our result indicates that LISA can distinguish the modified effect of LQG  with a parameter as small as $2\times10^{-6}$. The constraint on a parameter in LQG using the Fisher information matrix can be obtained within a fraction error of $10^{-6}$.

\end{titlepage}
}

\newpage

\section{Introduction}

General relativity (GR), as the most accurate gravitational theory, has passed a wide variety of tests in the Solar system \cite{Will:2014kxa} and binary pulsar \cite{Will:2014kxa, Wex:2014nva}, cosmological \cite{Clifton:2011jh, Ferreira:2019xrr}, as well as gravitational wave (GW) observations \cite{LIGOScientific:2019fpa, Baker:2017hug, LIGOScientific:2017ycc}. Although these successes align with the observational predictions of GR, there are other reasons to suggest that GR may not be a perfect theory of gravity. First, GR can not explain the dark matter and dark energy correctly, the accelerated expansion of the Universe \cite{SupernovaCosmologyProject:1998vns, SupernovaSearchTeam:1998fmf, Joyce:2014kja}, and the rotation curves of the galaxy. Second, it still does not solve the question of unifying GR and quantum mechanics. Third, the singularity problem in the event horizon can not be addressed where the physical laws are invalid. Motivated by solving these issues, kinds of alternatives to GR are proposed. Specifically, the modifications of GR at the quantum scale also have experienced several efforts; there are two promising candidate quantum gravity models, such as string theory and loop quantum gravity (LQG) \cite{Rovelli:1997yv, Addazi:2021xuf}.

To address the conflict between the classical theories of gravity and quantum gravity, the classical Big Bang and black hole singularities can be elegantly solved in the LQG, where a non-perturbative method in quantum gravity is resolved \cite{Modesto:2008im, Modesto:2009ve}. From LQG effective equations, some static, spherically symmetric, and non-rotating spacetimes have been constructed in \cite{Thiemann:2002nj, Gambini:2013ooa, Bodendorfer:2019jay, Bodendorfer:2019nvy, Bojowald:2018xxu, Assanioussi:2019twp, Gambini:2020nsf}, which are also called as the quantum extensions of the Schwarzschild black hole. From a non-spinning loop quantum gravity black hole (LQGBH) as a seed metric \cite{Bodendorfer:2019nvy, Bojowald:2018xxu},  one constructs a rotating spacetime using the Newman-Janis-Algorithm \cite{Newman:1965tw}, where the quantum effects described by a parameter could be rapidly decaying if an observer is away from the center and the exterior spacetime has a well-defined asymptotic region \cite{Brahma:2020eos}. Further, such spacetimes in binary black hole systems will help us in exploring both classical and quantum aspects of gravity due to their strong gravitational regimes, where the GR limits can be tested. While GR remains a classical theory, understanding how quantum effects could modify fundamental features of gravity may reveal important limitations of GR. In this line of endeavour, the LQG corrections result from non-perturbative approaches often stem from deeper theoretical principles \cite{Brahma:2020eos}, giving them the potential to alter classical dynamics in a manner that resolves the big bang singularity. However, a crucial question arises whether these quantum effects naturally diminish rapidly enough once beyond the Planck regime. Over the past almost 14 billion years of big bang, even the slightest quantum corrections could, in principle, accumulate and cause deviations from the predictions of GR over this large period \cite{Ashtekar:2011ni}. With this, from GW standpoint, if one perhaps collects large data for a longer inspiral duration for estimating observables through the phase evolution that accumulates over thousands of cycles, especially in the strong gravity regimes, one can possibly have observational consequences of such quantum signatures. This further motivates us to examine these effects through EMRIs, which may serve as a useful tool for low-frequency detectors like LISA.


Black holes are the most fascinating celestial bodies in the Universe, which are described by only three parameters: mass, angular momentum and charge according to no-hair theorem \cite{Misner:1963fr, Chrusciel:2012jk} in the context of GR. Testing Kerr nature of black holes has been conducted with dozens of GW events from coalescence of binaries compact objects observed by LIGO-Virgo-KAGRA 
\cite{LIGOScientific:2016aoc, Baker:2017hug, LIGOScientific:2017ycc, LIGOScientific:2019fpa, KAGRA:2021vkt}. It also becomes feasible to compute constraints on the fundamental theories in extreme environments of strong gravitational fields and large spacetime curvature \cite{LIGOScientific:2016dsl, LIGOScientific:2020tif, LIGOScientific:2021sio, Yunes:2024lzm}. These current tests do not support the additional charge of black holes predicted by the no-hair theorem, so it is essential to search for the possible deviation using the future GW detectors, such as the Einstein Telescope (ET) \cite{Punturo:2010zz}, Cosmic Explorer (CE) \cite{Dwyer:2014fpa}, LISA \cite{LISA:2017pwj}, TianQin \cite{TianQin:2015yph, TianQin:2020hid} and Taiji \cite{Ruan:2018tsw}. Because these detectors have the potential to detect many GW signals from more massive compact objects (COs) with a higher SNR, the future accessible GW events provide a valuable opportunity to accurately measure parameters of spacetime near the COs and place the rigorous constraint on GR and its alternatives \cite{Berti:2015itd, Jai-akson:2017ldo,Berry:2019wgg, LISA:2022kgy, Barausse:2020rsu}.

One of the target sources of LISA-like detectors is extreme mass-ratio inspirals (EMRI) with a smaller mass-ratio $\eta = \mu/M \in [10^{-7},10^{-4}]$, which contains a stellar-mass object and a massive black hole (MBH) located in the galactic center. The secondary body orbits about $\mathcal{O}(10^4-10^5)$ cycles around the MBH until it is captured at the last stable orbit (LSO); during the whole inspiralling, the GW signal emitted from the binary is precisely in the millihertz bank \cite{Berry:2019wgg, Berti:2019xgr}. Eccentric orbits in binary dynamics provide a more general framework. Studying EMRI with such orbits is important as they can start with significant nonzero eccentricity and can influence evolution of the system. The presence of eccentricity enables the orbits to exhibit radial velocity, which allows to probe a broader range of spacetime curvature, making estimates more sensitive to deviations from GR. Additionally, eccentricity enhances GW emission when both elements of the binary come closer, as well as it improves parameter estimation through the presence of both azimuthal and radial frequencies. This motivates our focus on eccentric dynamics to make a complete analysis. Further, EMRI observations could allow to measure parameters of MBH and test the fundamental characteristics of gravity with an unprecedented accuracy \cite{Barack:2006pq, Babak:2017tow, Fan:2020zhy, Zi:2021pdp}. Recent works based on EMRI broadly focus on studying the environmental effect surrounding MBH \cite{Kocsis:2011dr, Barausse:2014tra, Cardoso:2022whc, Dai:2023cft, Figueiredo:2023gas, Zi:2023omh, Rahman:2023sof, Rahman:2022fay, AbhishekChowdhuri:2023gvu}, test the nature of black hole horizon \cite{Sago:2021iku, Maggio:2021uge, Zhang:2021ojz,Ghosh:2024arw}, and detect the unique effects beyond GR, such as tidal heating \cite{Datta:2019euh, Datta:2019epe, Datta:2024vll}, tidal deformability \cite{Pani:2019cyc, DeLuca:2022xlz, Zi:2023pvl, Zhang:2024csc, AbhishekChowdhuri:2023gvu, Kumar:2024utz} and quantum effects \cite{Agullo:2020hxe, Datta:2021row, Fu:2024cfk,Chakravarti:2021jbv,Yang:2024lmj}. In this paper, we first consider computing the EMRI waveform from a spinning LQGBH, deriving the analytic expressions of orbital energy, angular momentum, and fundamental frequencies. Then, we assess the difference of EMRI waveforms between the Kerr BH and LQGBH by computing the dephasing and mismatch and show the constraint on LQGBH using the Fisher information matrix (FIM).

Our paper is organized as follows. In section (\ref{BHspacetime}),
we introduce the spinning LQGBH, and orbital motion equations and fundamental frequencies  in the subsection (\ref{gdscvecmtn}). The section (\ref{radiation:reaction}) presents the orbital energy and angular momentum loss due to the gravitational radiation reaction using the quadrupole formula. We show the method of computing waveform and data analysis in the section (\ref{wave}) and the results, including the waveform, dephasing, mismatch, and constraint on LQGBH in the section (\ref{result}). Finally, we show a brief summary of the results in section (\ref{discussion}).

 \par 
 \textit{Notation and Convention: } We set the fundamental constants $G$ and $c$ to unity and adopt the positive sign convention $(-1,1,1,1)$. Roman letters are used to denote spatial indices, and Greek letters are used to represent four-dimensional indices. 
 
\section{LQGBH Spacetime and orbital motion} \label{BHspacetime}
In this section, we briefly touch upon the description of a non-Kerr black hole spacetime that encapsulates a parameter which relates to quantum effects. We term such a spacetime rotating LQGBH, and the line element metric can be written in the Kerr-like form\cite{Brahma:2020eos, Afrin:2022ztr, Azreg-Ainou:2014pra}
\begin{align}\label{met}
ds^{2} = -\frac{\Psi}{\rho^{2}}\Big(\frac{\Delta}{\rho^{2}}(dt-a \sin^{2}\theta d\phi)^{2}-\frac{\rho^{2}}{\Delta}dr^{2}-\rho^{2}d\theta^{2}-\frac{\sin^{2}\theta}{\rho^{2}}[adt-(\omega(r)+a^{2})d\phi]^{2}\Big),
\end{align}
where $\rho^{2}=\omega(r)+a^{2}\cos^{2}\theta, \omega(r)=b^{2}, \Psi(r)=\rho^{2}$ and $\Delta=8L_{q}M^{2}\Tilde{a}b^{2}+a^{2}$. Further,
\begin{align}
b^{2}(x) = \frac{L_{1}}{\sqrt{1+x^{2}}}\frac{M^{2}+M^{2}(x+\sqrt{1+x^{2}})^{6}}{(x+\sqrt{1+x^{2}})^{3}} \hspace{3mm} ; \hspace{3mm} \Tilde{a}(x) = \frac{1+x^{2}}{b^{2}}\Big(1-\frac{1}{\sqrt{2L_{q}}}\frac{1}{\sqrt{1+x^{2}}}\Big),
\end{align}
where $x=\frac{r}{\sqrt{8L_{q}}M}\in (-\infty, \infty)$ and $L_{q}=\frac{1}{2}(l_{k}/MM_{W})^{2/3}\geq 0$. $L_{q}$ is a dimensionless parameter. ($M, M_{W}$) are the masses of the black hole and white hole, respectively \cite{Bodendorfer:2019nvy, Afrin:2022ztr}. We are interested in the case $M_W=M$, that is, a symmetric
bounce. By solving $\Delta(r)=0$, one can determine the location of the possible horizons. A numerical analysis of these solutions can be found in \cite{Brahma:2020eos, Afrin:2022ztr}.  Since obtaining complete analytical solutions to $\Delta(r)=0$, without any assumptions on the parameters ($L_q, a)$ is intractable, we take a linear-order approximation for deriving horizon whose resulting expressions are given by: $r_{\pm} = M\pm\sqrt{M^{2}-a^{2}}\mp\frac{4L_q M^{2}}{\sqrt{M^{2}-a^{2}}}$. It is evident $L_q=0$, gives Schwarzschild/Kerr horizons. The spacetime (\ref{met}) is also known as the LQG-inspired rotating black hole (LIRBH). Also, LIRBHs are regular everywhere and approach Kerr spacetime when quantum effects are absent ($L_q=0$). 

Furthermore, quantum effects emerge from the parameter $l_{k}$, appearing from holonomy modifications \cite{Bodendorfer_2019} and related to the parameter $L_{q}$. It is directly linked to the LQG theory's minimum area gap \cite{Brahma:2020eos}. We notice that the length scale $l_k$ goes as $l_{k}\propto M^{2}$. If we consider a small value $L_q = 2\times 10^{-6}$ \textemdash corresponding to the length scale of $\sqrt{l_{k}}\approx 132.078$ Km for a black hole of mass $10^{6}M_\odot$ \textemdash the position of the $r_{+}$ with $a=0.1$ exhibits only a minimal shift ($r_{+} \approx 2.95334242\times 10^{6}$ Km) similar to that in the Kerr geometry ($r_{+} \approx 2.95335423\times 10^{6}$ Km). Such deviations can translate into measurable differences in the waveforms, resulting from the phase evolution during the inspiral. Though these models provide singularity resolution of Kerr spacetime, the LIRBH metrics do not originate from a direct loop quantization of the Kerr geometry. Since LIRBHs are able to represent the effective regular spacetime description of LQG, they serve as potential sources in astrophysics \cite{Afrin:2022ztr}. Note that both the terminologies LQGBH and LIRBH are equivalent in the present article. Next, using spacetime geometry (\ref{met}), we construct orbital dynamics, including orbital frequencies. 

\subsection{Orbital motion}\label{gdscvecmtn}
This section provides necessary expressions and details of the eccentric orbital dynamics of the inspiralling object. The spacetime (\ref{met}) shows two integrals of motion ($E, J_{z}$). Since our analysis focuses on equatorial orbits, hence, the Carter constant $\mathcal{Q}=0$ \cite{Glampedakis:2002cb, Glampedakis:2002ya, PhysRevD.61.084004}. The geodesic velocities can be written in the following way, as described in (\ref{apenteu1}),

\begin{equation}
\begin{aligned}\label{gdscsnn}
\mu\Delta\rho\frac{dt}{d\tau}  =& \Big(\left(a^2+\omega\right) (a (a E -J_{z})+E \omega+a \Delta \left(J_{z}-a E \right)\Big) \\
\mu\Delta\rho\frac{d\phi}{d\tau} =& \Big(a (a (a E -J_{z})- E \Delta +E \omega +J_{z} \csc ^2\theta  \Delta \Big) \\
\mu^{2}\rho^{4} \Big(\frac{dr}{d\tau}\Big)^{2} =& \Big((a^{2}+\omega)E-a J_{z}\Big)^{2}-\Delta(\kappa+\mu^{2}\omega) 
\end{aligned}
\end{equation}
where with small black hole spin, $\kappa=J_{z}^{2}-2aEJ_{z}+\mathcal{O}(a^{2})$. $\kappa$ is the separation constant for radial and angular geodesic velocities that relates to the conventional Carter constant, $\mathcal{Q}\equiv \kappa-(J_{z}-a E )^{2}$, which is zero for the case of equatorial consideration. The expressions in Eq. (\ref{gdscsnn}) are the velocities of the inspiralling object (timelike geodesic velocities). The quantities $(E, J_z)$ are orbital energy per unit mass (mass of point particle $\mu$), orbital angular momentum per unit mass along $z$-direction. Thus ($E, J_z$) are two constants of motion. We will use these velocities for the computation of the rate of change of energy and angular momentum under the assumption of small black hole spin, i.e., in the linear order $\mathcal{O}(a)$, as also discussed in the appendix (\ref{apenteu1}). Note that we perform our calculation by making the quantities dimensionless with respect to the central black hole mass $M$ \cite{Kumar:2024utz}, for instance $\hat{r}=r/M, \hat{E}=E/\mu, \hat{J}_z=J_z/(\mu M), \hat{a}=a/M$; where quantities denoted with a hat are dimensionless. However, for notational convenience, we do not adopt any specific notation for such dimensionless quantities throughout the article. One can always express these in physical units at a later stage.

\begin{figure}[h!]\centering
\includegraphics[width=3.0in, height=2.1in]{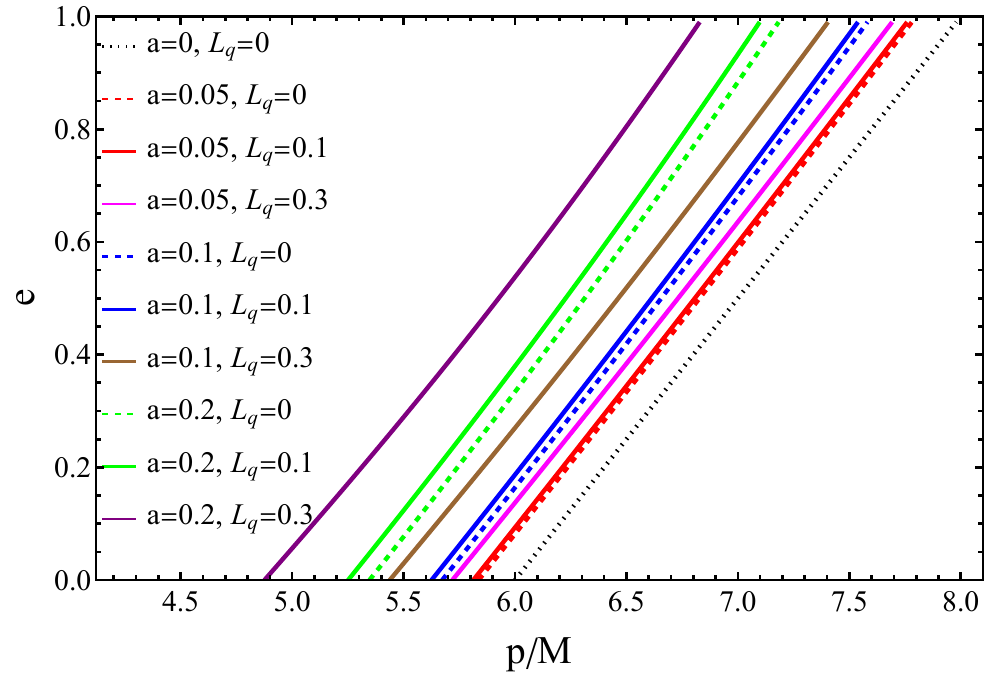}
\caption{The plot shows separatrix curves in the ($p, e$) plane for various ($a, L_q$). The dotted line is Schwarzschild ($p = 6 + 2e$); dashed lines are Kerr ($L_q=0$). Bounded orbits lie right of the separatrix, which marks the LSO boundary.} \label{sep}
\end{figure}

The radial expression in Eq. (\ref{gdscsnn}) governs the orbital motion on the equatorial plan. Following \cite{AbhishekChowdhuri:2023gvu, Kumar:2024utz, Rahman:2023sof}, and noticing the fact that eccentric orbits will possess two turning points: \textit{periastron} ($r_{p}$) and \textit{apastron} ($r_{a}$): $r_{p}=p/(1+e)$ and $r_{a}=p/(1-e)$. The bounded orbits can be obtained in the range $r_{p}<r<r_{a}$ if $V_{eff}(r)<0\,.$ This is maintained only when $V'_{eff}(r_{a})>0$ and $V'_{eff}(r_{p})\leq 0$ \cite{Rahman:2023sof, AbhishekChowdhuri:2023gvu}, where prime denotes the derivative with respect to $r$. Since the radial velocity vanishes at the turning points; hence, $V_{eff}\vert_{r=r_{p},r_{a}}=0$. We obtain the integrals of motion ($E, J_{z}$) that show the impact of $L_{q}$. 

\begin{equation}
\begin{aligned}\label{cnst1}
E =& \frac{1}{p}\sqrt{\frac{-4(1-e^{2})^{2}L_{q}+p((p-2)^{2}-4e^{2})}{p-3-e^{2}}} - \frac{a(1-e^{2})^{2}}{p^{2}(p-3-e^{2})}\sqrt{\frac{p^{2}+2L_{q}(3+e^{2}-3p)}{p-3-e^{2}}}+\mathcal{O}(a^{2})\\
J_{z} =& \sqrt{\frac{p^{2}+2L_{q}(3+e^{2}-3p)}{p-3-e^{2}}}-\frac{a(3+e^{2})}{p(p-3-e^{2})}\sqrt{\frac{-4(1-e^{2})^{2}L_{q}+p((p-2)^{2}-4e^{2})}{p-3-e^{2}}}+\mathcal{O}(a^{2})
\end{aligned}
\end{equation}
Further, we determine the location of the last stable orbit (LSO) with the use of Eq. (\ref{cnst1}) \cite{Glampedakis:2002cb, PhysRevD.50.3816, PhysRevD.77.124050, AbhishekChowdhuri:2023gvu}. It is also termed \textit{separatrix}. It gives the area for ($p, e$) that divides the bound and unbound orbits. Additionally, it provides the lowest semi-latus rectum value that is permitted for all bound orbits ending for a given $e$. The corresponding expression has the following form:
\begin{equation}
\begin{aligned}\label{sp1}
p_{sp}=& 2(e+3)-\frac{a\sqrt{(4(3+e^{2})^{2}+p_{1})(8(1+e)(3+e)-p_{2})}}{(3+e)^{2}+p_{3}}+\mathcal{O}(a^{2})
\end{aligned}
\end{equation}
where $p_{1}=2 (-15 + (-6 + e) e) L_{q}, p_{2}=(1 - e^2)^2 L_{q}$ and $p_{3}=(-15 + e (-8 + 3 e)) L_{q}$. If we switch off parameter $L_{q}$, $p_{sp}$ reduces to the Kerr case. The results comply with \cite{PhysRevD.50.3816, Glampedakis:2002ya, AbhishekChowdhuri:2023gvu, Kumar:2024utz} and provide us with the truncation region of the inspiralling object. However, for small values of $L_q= 2\times 10^{-6}$ (equivalent to the length scale $\sqrt{l_{k}}\approx 132.078$ Km for $M=10^{6}M_\odot$), the separatrix does not lead to a large deviation and remains at the order of $\sim 10^{6}$ Km, the same as in the case of the Kerr/Schwarzschild. Nevertheless, as it will be shown in the subsequent section that these seemingly small deviations can lead to measurable differences in waveform characteristics, particularly in the mismatch, which can possibly be detected by future space-based low-frequency detectors. Moreover, Fig. (\ref{sep}) presents the nature of LSO or separatrix, with a comparison of Schwarzschild and Kerr cases, for distinct values of $L_q$. It indicates that as we increase $L_q$, the separatrix has the larger deviation from the case when $L_q=0$. 

As mentioned earlier, the eccentric dynamics will have the bounded orbits in the range ($r_{p}, r_{a}$). Since, there are diverging behaviour in differential equations at the turning points while proceeding with orbital dynamics, we introduce a radial parametrization that helps overcome such an issue \cite{PhysRevD.50.3816}. The parameterization is 
\begin{align}\label{prmtz}
r=\frac{p}{1+e\cos\chi}\,.
\end{align}
These points ($r_{p}, r_{a}$) correspond to ($\chi=\pi, \chi=0$), respectively.
Using the parameterization scheme \eqref{prmtz}, the radial and azimuthal frequencies  for the geodesic eccentric orbits on the  equatorial plane  can be 
obtained with
\begin{eqnarray}
\Omega_r = \frac{2\pi}{T_r}, \quad \Omega_\phi = \frac{\Delta \phi}{T_r}, 
\end{eqnarray}
where $T_r=\int_0^{2\pi} \frac{dt}{d\chi} d\chi $ is the radial period and $\Delta \phi = \int_0^{2\pi} \frac{d\phi}{d\chi} d\chi $ is the period in the azimuthal direction.
The expressions for the radial and azimuthal frequency can be given by

\begin{eqnarray}
T_r & = &2\pi (p^{5/2} - 6(1-e^2)L_q - 2(1-e^2)L_q \sqrt{p(1-e^2)})/(p(1-e^2)^{3/2})  \nonumber  \\ &+& 2\pi (3a(-1+e^2)-3a(-1+e^2)(9+2\sqrt{1-e^2})L_q)/(p(1-e^2)^{3/2}),  \\ \nonumber 
\Omega_\phi & = & a(e^2-1)(\sqrt{1-e^2}(1+3e^2)p -6L_q(-1-e^4+2\sqrt{1-e^2} + e^2(2+6\sqrt{1-e^2})))/p^4  \nonumber  \\ &+&
\sqrt{1-e^2}(-1+e^2)( -2p^{5/2}+L_q(4\sqrt{L_q}(4+5e^2) +6p^{3/2} +4(e^2-1)\sqrt{p(1-e^2)} ) +6\sqrt{p}L_q^2) /(2p^4), \nonumber   \\
\end{eqnarray}
One can then determine the GW dephasing $\delta\Phi_{\phi,r}$ by estimating the phase evolution $\Phi(t)$ for both GR and additional effects that can be computed by integrating the difference in frequency evolution. In other words, we can compute the difference of  orbital phases between the Kerr and LQGBH cases by defining dephasing
\begin{eqnarray}
\delta \Phi_{\phi,r} = 2\int_0^t \left(\Omega_{\phi,r}^{L_q=0}-\Omega_{\phi,r}^{L_q\neq0}\right)
\end{eqnarray}
where $\Omega_{\phi,r}^{L_q=0}$ is the orbital frequencies in the Kerr spacetime 
and $\Omega_{\phi,r}^{L_q\neq0}$ denotes the orbital frequencies in the LQGBH spacetime.
In general, the threshold value  of dephasing is set as $\delta \Phi \sim 1$ rad \cite{Datta:2019epe,Zi:2023qfk,Zi:2023omh}, the emission of LQGBH may be discerned by LISA if the dephasing is larger than 1 rad. It should be noted that the criterion is a rule of thumb, which needs be validated with a careful analysis.


\section{Fluxes and orbital evolution} \label{radiation:reaction}

In this section, we study the effect of radiation reaction on the inspiralling object, and, as a consequence, we estimate the rate of change of orbital energy and angular momentum as well as the eccentric orbital evolution that records the imprints of deformations. The radiation reaction causes the integrals of motion ($E, J_z$) to evolve over time \cite{Hughes:2024tja}. Following \cite{Flanagan:2007tv, Ryan:1995xi, Misner:1973prb, Mukherjee:2022pwd}, also discussed in appendix (\ref{apenteu1}), we separate out the rest mass energy as $E=\mu +\mathcal{E}$. Next, we investigate the impact of quantum effects with the leading-order PN analysis \cite{Flanagan:2007tv, PhysRevD.52.R3159, Ryan:1995xi} within the equatorial consideration. 

We start by re-stating the useful expressions mentioned in the appendix in terms of Cartesian coordinates ($x_{1},x_{2},x_{3})=(r\sin\theta \cos\phi, r\sin\theta \sin\phi, r\cos\theta$). 
\begin{equation}
\begin{aligned}\label{re}
\mathcal{E} =& \frac{1}{2}\dot{x}_{i}\dot{x}_{i}-\frac{1}{\sqrt{x_{i}x_{i}}}+\frac{1}{x_{i}x_{i}}(4+3 L_{q}) \\
J_{z} =& \epsilon_{3jk}x_{j}\dot{x}_{k}\left(1+\frac{2L_{q}}{r^{2}}+\frac{4L_{q}^{2}}{r^{4}}+\frac{8L_{q}^{3}}{r^{6}} \right)
\end{aligned}
\end{equation}
where $r^{2}\sin^{2}\theta \dot{\phi}=\epsilon_{3jk}x_{j}\dot{x}_{k}$. Note that terms proportional to velocity in Eq. (\ref{re}) will contribute to the energy and angular momentum loss computation of the orbit because it eventually relates to radiation reaction acceleration ($a_{j}$ or $\Ddot{x}_{i}$). The dot signifies differentiation with respect to the coordinate time $t$. The expressions comply with \cite{Flanagan:2007tv, AbhishekChowdhuri:2023gvu} when taken $L_{q}\rightarrow 0$. 

Due to the radiation reaction effect, the instantaneous fluxes can be written by using Eq. (\ref{re})
\begin{align}\label{inst fluxes}
\dot{\mathcal{E}} = x_{i}\Ddot{x}_{i} \hspace{5mm} ; \hspace{5mm}
\dot{J}_{z} = \epsilon_{3jk}x_{j}\Ddot{x}_{k}\left(1+\frac{2L_{q}}{(x_{i}x_{i})}+\frac{4L_{q}^{2}}{(x_{i}x_{i})^{2}}+\frac{8L_{q}^{3}}{(x_{i}x_{i})^{3}} \right). 
\end{align}
Again, we ignore the velocity-independent terms, as those will not contribute to calculating the loss of energy and angular momentum of the inspiralling object. The acceleration term ($\Ddot{x}_{i}$) is often denoted as $a_{j}$ and given by  \cite{Flanagan:2007tv, PhysRevD.52.R3159}
\begin{align}\label{accelration}
a_j=-\dfrac{2}{5}I^{(5)}_{jk}x_{k}+\dfrac{16}{45}\epsilon_{jpq}J^{(6)}_{pk}x_{q}x_{k}+\dfrac{32}{45}\epsilon_{jpq}J^{(5)}_{pk}x_{k} \dot{x}_{q}+\dfrac{32}{45}\epsilon_{pq[j}J^{(5)}_{k]p}x_{q} \dot{x}_{k}+\dfrac{8J}{15}J^{(5)}_{3i},
\end{align}
where $B_{[ij]}$ is an anti-symmetric quantity, $B_{[ij]}=\frac{1}{2}(B_{ij}-B_{ji})$. The superscripts are the derivative orders. $J$, the last term, denotes black hole spin $a$. We notice that the radiation reaction acceleration depends on two symmetric trace-free (STF) quantities known as mass ($I_{jk}$) and current ($J_{jk}$) quadrupole moments, given by
\begin{equation}\label{moments}
I_{jk}=\Big[x_j x_k\Big]^{\text{STF}} \hspace{3mm} ; \hspace{3mm}
J_{jk}=\Big[x_{j}\epsilon_{kpq}x_{p}\dot{x}_{q}-\dfrac{3}{2}x_jJ\delta_{k3}\Big]^{\text{STF}}.
\end{equation}
We further utilize the expression of radial velocity mentioned in Eq. (\ref{gdscs3n}), and we get,
\begin{equation}
\begin{aligned}\label{enrjz2}
\mathcal{E} =& \frac{e^2-1}{2 p}+\left(e^2-1\right)^2 \left(\frac{9 a L_{q}}{p^{7/2}}-\frac{a}{p^{5/2}}+\frac{12 L_{q}^2}{p^4}-\frac{2 L_{q}}{p^3}\right) \\
J_{z} =& \sqrt{p}-\frac{a \left(e^2+3\right)}{p}+\left(e^2+1\right) \left(\frac{12 a L_{q}}{p^2}+\frac{12 L_{q}^2}{p^{5/2}}-\frac{4 L_{q}}{p^{3/2}}\right)-\frac{9 L_{q}^2}{2 p^{3/2}}-\frac{3 L_{q}}{\sqrt{p}}
\end{aligned}
\end{equation}
Here, we have presented the higher-order terms in $p$ to see how the nature of parameter $L_{q}$ goes. This sums up the full setup to obtain the instantaneous loss in energy and angular momentum in terms of ($p, e, \chi$). Then we average over $\chi\in (0, 2\pi)$. The averaged quantities, denoted as ($<\Dot{\mathcal{E}}>, <\Dot{J_{z}}>$), over the course of a single orbit can be written as
\begin{align}
<\Dot{\mathcal{E}}> = \frac{1}{T_{r}}\int_{0}^{2\pi}\Dot{\mathcal{E}}\frac{dt}{d\chi}d\chi \hspace{3mm} ; \hspace{3mm} <\Dot{J_{z}}> = \frac{1}{T_{r}}\int_{0}^{2\pi}\Dot{J_{z}}\frac{dt}{d\chi}d\chi.
\end{align}
As a result, the average loss of energy and angular momentum is
\begin{equation}
\begin{aligned}\label{avflxn}
<\Dot{\mathcal{E}}> =& -\frac{(1-e^{2})^{3/2}}{5p^{5}}\Big[\frac{1}{3}\left(37 e^4+292 e^2+96\right)-\frac{24 \left(33 e^4+104 e^2+24\right) L_{q}}{p}\Big] \\
<\Dot{J}_{z}> =& -\frac{4(1-e^{2})^{3/2}}{5p^{7/2}}\Big[\left(7 e^2+8\right)-\frac{3 \left(2 e^4+63 e^2+40\right) L_{q}}{p}\Big].
\end{aligned}
\end{equation}
Note that we have discarded subleading terms here and presented higher order terms in appendix (\ref{apenf}). This presents the leading order correction of the parameter $L_{q}$ that indicates the quantum effects in observables. The above losses are written up to 1PN order. Further, one can restore the power of speed of light (c) to track the PN order. For the PN counting, the expressions go as: $<\dot{\mathcal{E}}> \sim \frac{1}{p^{5}c^{5}}\Big(a_{1}+\frac{a_{2}L_{q}}{pc^{2}}\Big)$ and $<\dot{J}_{z}>\sim \frac{1}{p^{7/2}c^{5}}\Big(a_{3}+\frac{a_{4}L_{q}}{pc^{2}}\Big)$. Where ($a_{1}, a_{2}, a_{3}, a_{4}$) are functions of eccentricity ($e$) which can be seen in the expressions (\ref{avflxn}). This implies that the $L_{q}$, emerging at 1PN, has a dominant effect than the spin ($a$), which appears at 1.5PN. Note that we have not considered the small $L_{q}$ assumption; the Eq. (\ref{avflxn}) reflects the emergence of $L_{q}$ upon the expansion of semi-latus rectum ($p$) about infinity. One may consider the higher-order expansion in $p$. It is to add that at 1.5PN, we do not find the cross contributions such as $\mathcal{O}(a L_{q})$. The non-linear nature of $L_{q}$ appears at the higher-order, which we have mentioned in the appendix (\ref{apenf}). Also, if we switch off the parameter $L_{q}$, the results are consistent with the existing literature \cite{AbhishekChowdhuri:2023gvu, Flanagan:2007tv, Glampedakis:2002ya, Ryan:1995xi, PhysRev.131.435, PhysRev.136.B1224}. 

Next, we determine the orbital evolution of the inspiralling object. Following \cite{AbhishekChowdhuri:2023gvu, Kumar:2024utz},
\begin{equation}
\begin{aligned}
\Big\langle\frac{dp}{dt}\Big\rangle = \Big(\frac{\dot{\mathcal{E}}\partial_{e}J_{z} -\dot{J}_{z}\partial_{e}\mathcal{E}}{\partial_{p}\mathcal{E}\partial_{e}J_{z}-\partial_{e}\mathcal{E}\partial_{p}J_{z}}\Big) \hspace{3mm} ; \hspace{3mm} \Big\langle\frac{de}{dt}\Big\rangle = \Big(\frac{\dot{J}_{z}\partial_{p}\mathcal{E}-\dot{\mathcal{E}}\partial_{p}J_{z}}{\partial_{p}\mathcal{E}\partial_{e}J_{z}-\partial_{e}\mathcal{E}\partial_{p}J_{z}}\Big)\,.
\end{aligned}
\end{equation}
We obtain,

\begin{equation}
\begin{aligned}\label{dpdtn}
\Big\langle\frac{dp}{dt}\Big\rangle =& -\frac{8(1-e^{2})^{3/2}}{5p^{3}}\Big((8+7e^{2})-\frac{6  \left(e^4+35 e^2+24\right) L_{q}}{p}\Big) \\
\Big\langle\frac{de}{dt}\Big\rangle =& -e\frac{(1-e^2)^{3/2}}{5p^4}\Big(\frac{1}{3}(304+121e^{2})-\frac{24 \left(e^4+67 e^2+93\right) L_{q}}{p}\Big)
\end{aligned}
\end{equation}
Again, the subleading terms in $p$ have been discarded here and presented in the appendix (\ref{apenf}). One can solve the Eq. (\ref{dpdtn}) simultaneously for obtaining ($p(t), e(t)$), which will show the impact of parameter $L_q$ in the orbital evolution of the inspiralling object. Further, one can also show the amount of time taken by the inspiralling object to reach the LSO. In particular, we are interested in examining the effect of $L_q$ on the evolution timescale. We obtain such a quantity by subtracting the GR part. We can consider the inspiral starts at $p=16$ and reaches the separatrix at $p=p_{sp}$. Now, using Eq. (\ref{dpdtn}), we can integrate with respect to $p$. As a result, we arrive at
\begin{align}
\Delta t \approx -\frac{5(24+35e^{2}+e^{4})(-2744+p_{sp}^{3})}{4(1-e^{2})^{3/2}(8+7e^{2})^{2}}L_q.
\end{align}
This analysis illustrates the shift in the timescale needed for the inspiralling object to reach the LSO. It is worth mentioning that we have taken $\mathcal{O}(L_{q})$ order term only for computing such a quantity in order to see the impact of $L_q$. Consequently, we notice if the $\Delta t$ is negative, it indicates that the secondary takes less time to reach the LSO when the $L_q$ is turned on. This analysis is particularly independent of computations carried out in the next section and onwards. Let us now touch upon the methodology for generating waveform and detectability of the parameter $L_q$.  


\section{Waveform and data analysis}\label{wave}
In this section, we show the quadruple formula of the waveform from the EMRI system and GW data analysis method. The GW waveform signals will be affected by the presence of the parameter $L_{q}$. As a result, GWs will exhibit noticeable changes, which can be measured through space-based detectors.

When the inspiral geodesic orbits are obtained, the quadruple approximation formula for the EMRI waveform can be written as follows in the transverse-traceless (TT) gauge
\begin{equation}
h_{ij}^{\rm TT} = \frac{2}{D} \left(P_{il}P_{jm} -\frac{1}{2}P_{ij} P_{lm}\right) \Ddot{I}_{lm}
\end{equation}
where $D$ is the distance from source to detector, $P_{ij} = \delta_{ij}-n_i n_j$ is the projection operator into the unit drection $n_{ij}$ of wave and $\delta_{ij}$ is the Kronecker delta symbol. Then, two polarizations for the plus and cross can be given by
\begin{eqnarray}
h_+ &=&-(\Ddot{I}_{11} - \Ddot{I}_{22})(1+\cos\iota) \\
h_\times &=& \mathcal{A} \sin(2\Phi(t)+2\Phi_0)\cos\iota
\end{eqnarray}
where $\mathcal{A} =2m (M \omega(t))^{3/2} /D$, $\iota$ denotes the inclination angle
between the orbtital angular momentum and the detector.
The mass quadrupole moment for binary objects is obtained with the stress-energy tensor
\begin{equation}
I_{ij} = \int d^3x T(t,x^i)x^ix^j=m z^i(t)z^j(t)
\end{equation}
where $ T(t,x^i) = m \delta^{3} (x^i-z^i(t))$ is the stress-energy tensor of source and  $z^i$ is the worldline of point particle  in the  Cartesian coordinates  $x^i$.
At last, the GW signal responded by the LISA-like  detectors in the low-frequency can be written as  \cite{Apostolatos:1994mx,Cutler:1997ta}
\begin{eqnarray}\label{eq:hphc}
h^{\rm I,II}(t) = \frac{\sqrt{3}}{2}[h_+(t)F^{\rm I,II}_+(t) + h_\times(t)F^{\rm I,II}_\times(t)]
\end{eqnarray}
where $F^{\rm I,II}_{+,\times}(t)$ are the interferometer pattern function for the space-borne detectors,
the full expression depends on the source orientation $(\theta_S,\phi_S)$,
and the direction of MBH spin $(\theta_K,\phi_K)$ in the ecliptic coordinate \cite{Apostolatos:1994mx,Cutler:1997ta}.

To distinguish the difference of the EMRI waveforms from the Kerr spacetime and
LQGBH spacetime, we can obtain the mismatch via defining the overlap
between two waveforms $h_a$ and $h_b$
\begin{eqnarray}\label{eq:mismatch}
\mathcal{M} \equiv 1 - \mathcal{O}(h_a|h_b),
\end{eqnarray}
where the overlap $\mathcal{O}(h_a|h_b)$ is given by inner product \cite{Cutler:1997ta}

\begin{equation}
\mathcal{O}(h_a|h_b) = \frac{<h_a|h_b>}{\sqrt{<h_a|h_a><h_b|h_b>}},
\end{equation}
and the noise-weighted inner product $<h_a|h_b>$ is defined by
\begin{align}\label{inner}
<h_a |h_b > =2\int^{f_b}_{f_a} df \frac{h_a^*(f)\tilde{h_b}(f)+\tilde{h_a}(f)h_b^*(f)}{S_n(f)}.
\end{align}
Here the tilde and star stand for the Fourier transform and complex conjugation,
the noise  power spectral density $S_n(f)$ is the space-borne GW detector LISA \cite{LISA:2017pwj} and the range of integration is set as $f_a = 10^{-4}$ Hz, $f_b = 1$ Hz.
Note that the mismatch $\mathcal{M}=0$ when two waveforms are identical. Consequently, the overlap becomes $\mathcal{O}=1$.
An empirical formula for distinguishing two kinds of GW signals detected by LISA is proposed, where one can claim that the detector can discern one waveform 
$h_a$ from the other waveform $h_b$ if the mismatch satisfies the inequality 
$\mathcal{M}\geq 1/(2\rho^2)$ \cite{Flanagan:1997kp,Lindblom:2008cm}.
Generally, the signal-to-noise ratio (SNR) of the EMRI signal in the source parameter evaluation for the LISA or TianQin is a moderate value $\rho = 20$ \cite{Babak:2017tow, Fan:2020zhy}, so the threshold value of mismatch distinguished by LISA is $\mathcal{M}=0.00125$.

Finally, we compute the FIM to extract the measurement error of source parameters \cite{Vallisneri:2007ev}.
If the GW signal has a higher SNR $\rho$, the variance-covariance matrix can be written as
\begin{eqnarray}
<\delta \lambda^i \delta \lambda^j  >=  (\Gamma^{-1})^{ij} \left(1+\mathcal{O}(1/\rho)\right) \simeq   \Sigma^{ij} ,
\end{eqnarray}
then the uncertainties of source parameters
can be given by
\begin{eqnarray}
\Delta \lambda^i = \sqrt{<(\delta\lambda^i)^2>} \simeq  \sqrt{(\Gamma^{-1})^{ij}} ,
\end{eqnarray}
where $<>$ denotes  the expectation value related with the noise \cite{Canizares:2012is}, 
the physical meanings of source parameters $ \lambda^i $ are listed in Table \ref{source:parameters} and
the definition of FIM $\Gamma$ can be written as follows:
\begin{equation}
\Gamma^{ij}=\Big< \frac{\partial h}{\partial \lambda^i} \Big| \frac{\partial h}{\partial \lambda^j} \Big>.
\end{equation}
with $\lambda^i$, $i=1, 2,...,$ are the parameters appearing in the waveform, the full description
can see Table \ref{source:parameters}, and the inner product $<|>$ is  defined by Eq. \eqref{inner}.
Note that when the linear signal approximation is satisfied, the FIM method is applicable, which has been discussed in Refs. \cite{Vallisneri:2007ev,Maselli:2021men,Zi:2022hcc}.
The numerical stability of the inverse FIM is discussed in Appendix \ref{appB:Stability}.
Let us now analyze the results and constrain $L_q$ with LISA observations.

Finally, we estimate the measurement error of solid angles using the combination of the bound of angles $(\theta^{}_{\rm S},  \phi^{}_{\rm S},  \theta^{}_{\rm K},\phi^{}_{\rm K})$
\begin{equation}
\Delta \Omega_i = 2\pi |\sin\theta_i|\sqrt{\Sigma^2_{\theta_i}\Sigma^2_{\phi_i}-\Sigma^2_{\theta_i\phi_i}},
\end{equation}
where $i\in\{S,K\}$.

\begin{table}
\caption{Physical meanings of the source parameters for EMRI system with an LQGBH.
The subindex $0$ denotes the parameter given at the initial time.
Four angles $(\theta_S, \phi_S)$ and $(\theta_K, \phi_K)$ are the spherical polar coordinates with respect to the ecliptic, which describe
the direction of EMRI and MBH's spin. The luminosity distance is set to $D = 1\,$Gpc.  \label{source:parameters}}
\begin{tabular}{ll}
\hline
Parameters                                    & Physical meanings \\
\hline
$M$                                         & mass of MBH. \\
$a$                                         & spin of MBH. \\
$m$                                         & mass of secondary body. \\
$e_0$                                       & eccentricity of the orbit at $t_0$. \\
$p_0$                                       & dimensionless semilatus rectum at $t_0$. \\
$L_q$                                       & dimensionless parameter. \\
$\theta_{\rm S}$                            & polar angle of source. \\
$\phi_{\rm S}$                              & azimuthal angle of source. \\
$\theta_{\rm K}$                            & polar angle of MBH spin. \\
$\phi_{\rm K}$                              & azimuthal angle of MBH spin. \\
$D$                                         & distance from the detector to the source. \\
$\chi_0$                                    & angle variable for the radial motion. \\
$\Phi_0$                                    & azimuthal angle in Boyer-Lindquist coordinate. \\
\hline
\end{tabular}
\end{table}

\section{Results and detectability}\label{result}
In this section we present some results regarding the EMRI waveform modified by LQGBH, dephasing and mismatch to assess the difference of orbital phase and constraint on LQGBH with space-based detectors.

\begin{figure*}[!h]
\includegraphics[width=16cm, height =7.5cm]{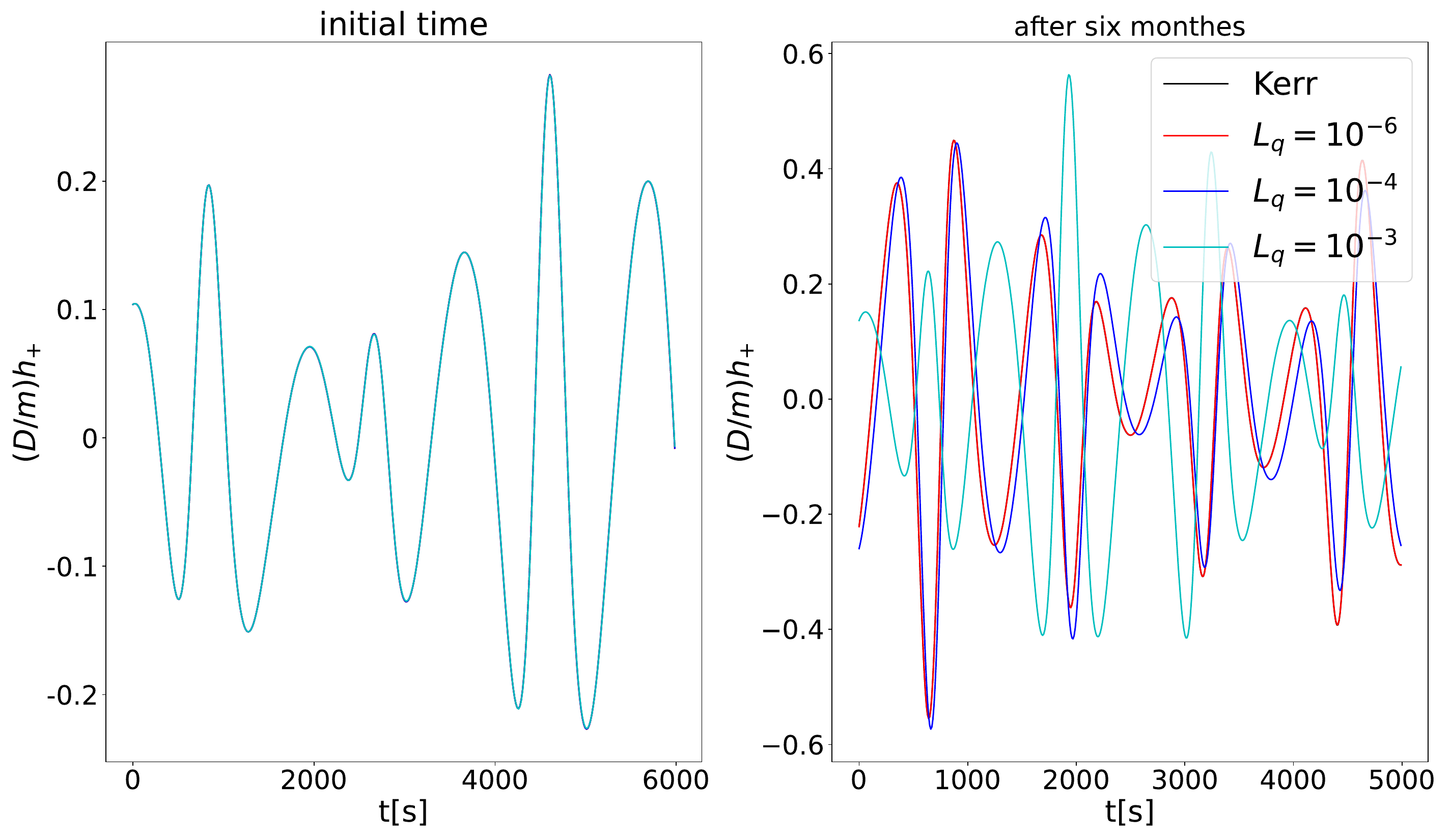}
\caption{Comparison between the polarizations $h_+$ of four EMRI waveforms for the standard Kerr and LQGBHs with a  $L_q \in\{10^{-6},10^{-4},10^{-3}\}$ cases.
The spin of LQGBH  and the initial orbital parameters
are set to $a=0.1$, $p_0=16$ and $e_0=0.5$.
The left panel is the initial stage of the time domain waveforms
and the right panel denotes the time domain waveforms after six months.
The inspiral time of CO is set as four months, the displaying results are only plotted for one day to show concisely.
}
\label{fig:wave}
\end{figure*}
\begin{figure*}[ht!]
\includegraphics[width=8.5cm, height =6.0cm]{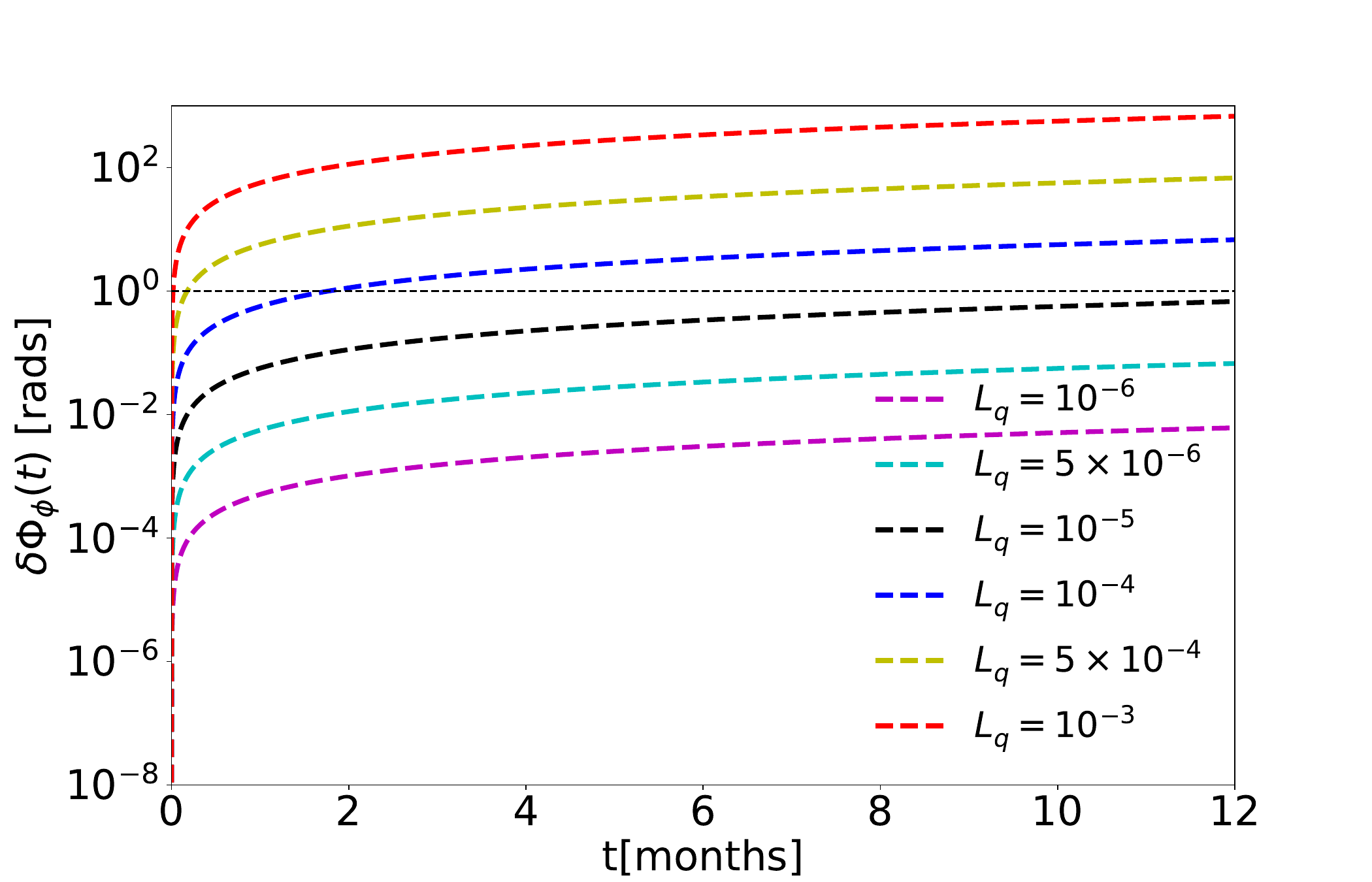}
\includegraphics[width=8.5cm, height =6.0cm]{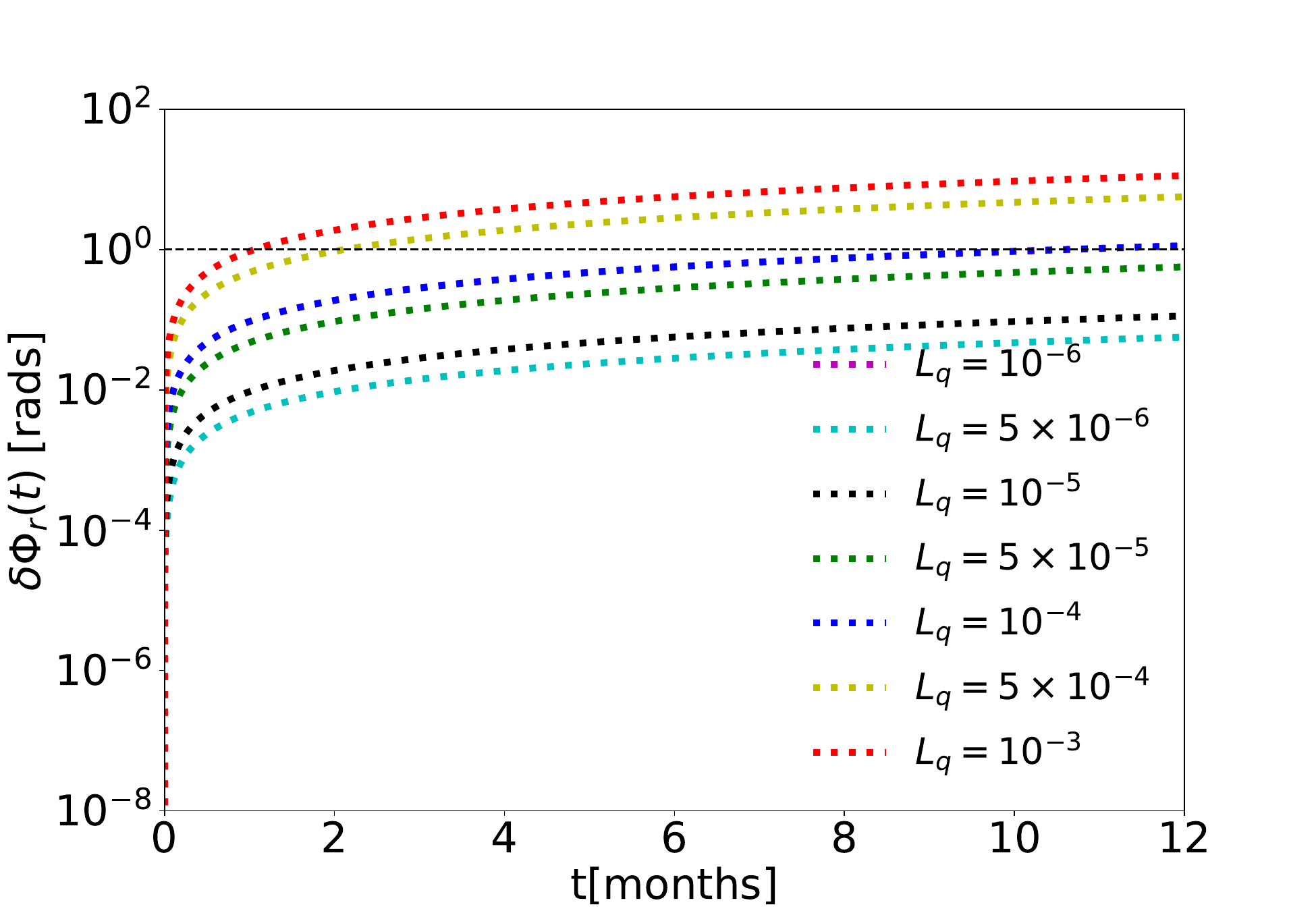}
\caption{Azimuthal (left) and radial (right) dephasings are as a function of the observation time, where the spin of LQGBH  $a=0.1$ and the dimensionless parameter $L_q \in\{10^{-6},5\times10^{-6},10^{-5},5\times10^{-5},10^{-4},5\times10^{-4},10^{-3}\}$. The initial orbital semi-latus rectum and eccentricity are set to $p_0=16$ and $e_0=0.5$. The horizontal black dashed line in the figures denotes the threshold for the phase that can be distinguished by LISA.
}\label{fig:dephasing}
\end{figure*}
First, we present the waveform in the time domain at the initial stages, where the inspiral persists for 5000 seconds after six months, as shown in Fig. \ref{fig:wave}. It is evident that the four waveforms are nearly identical during the first 6000 seconds; however, a noticeable phase difference is increasingly apparent to the naked eye, even when the parameter $L_q$ slightly increases, leading to more pronounced results from the detection perspectives of such a parameter. Second, Fig. \ref{fig:dephasing}  shows the radial and azimuthal dephasings as a function of observation time, where the spin of MBH is set to $a=0.1$ and
the inspiral time of secondary CO is one year.
On the whole, the accumulation of the radial and azimuthal dephasings are both gradually increasing, ranging from a fraction  to $10^2$ rad.
The horizontal black dashed line is threshold value $\delta \Phi_{r,\phi}=1$ rad that is distinguished by LISA, beyond which the modified effect of LQGBH can potentially be resolvable by LISA. Overall, the dephasings for two motions are all bigger with the
quantum-corrected parameter $L_q$ is larger.
Specifically, for the azimuthal dephasing $\delta \Phi_\phi$ in the left panel, after the observation of two months, the inspirals can result in a dephasing larger than the threshold for the fixed quantum-corrected parameter $L_q=10^{-4}$. However, one year observation of LISA do not reach to a recognizable 
threshold; it may be possible to be resolvable if the observation time increases slightly. For the right panel, after one year of observation, LISA can just resolve the effect of LQGBH for the fixed same parameter $L_q=10^{-4}$.
Therefore, the azimuthal dephasing always is bigger than the radial case, which is  also consistent with \cite{Barsanti:2022ana,Zi:2023qfk}.

\begin{figure*}[ht]
\includegraphics[width=8.5cm, height =6.0cm]{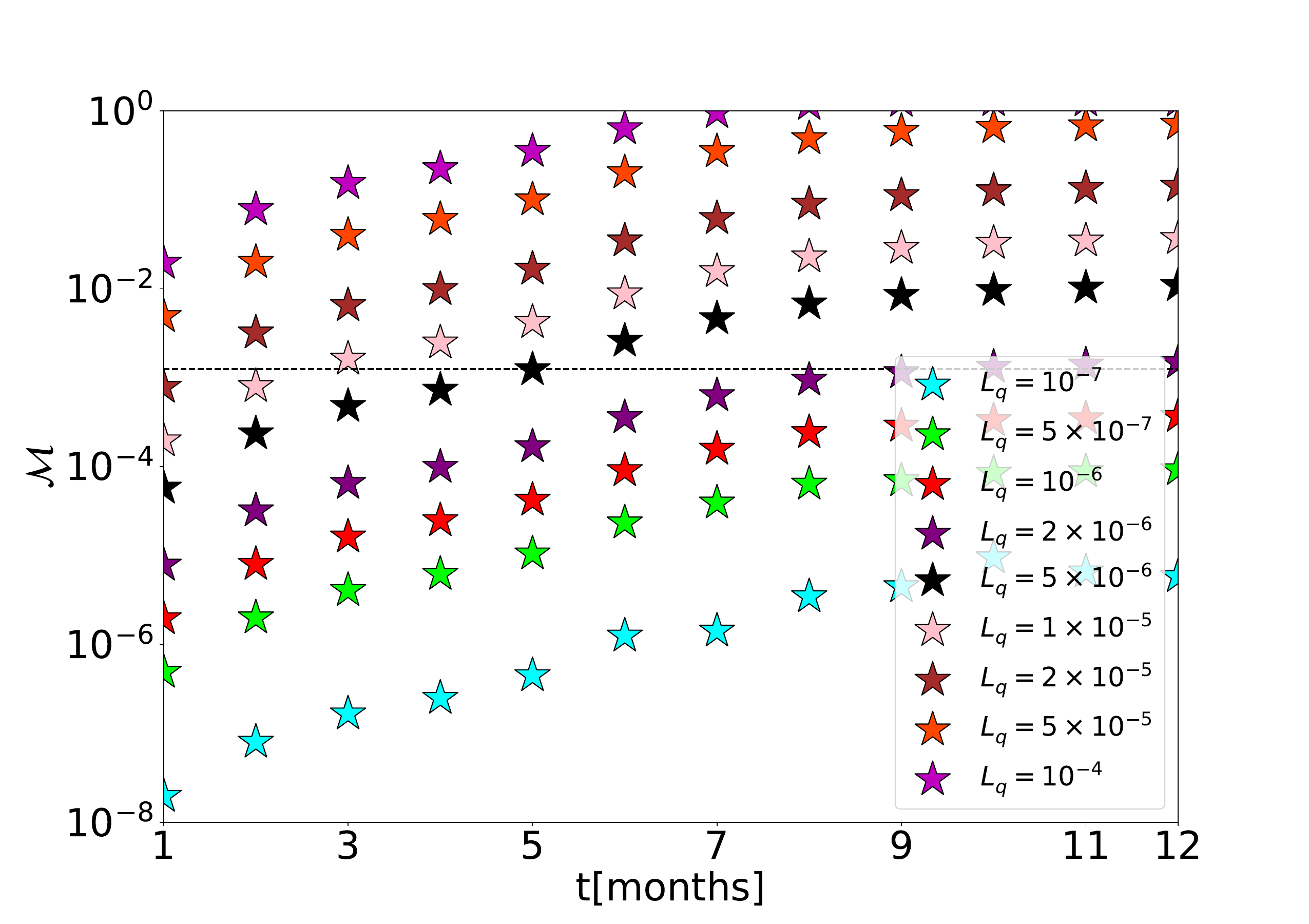}
\includegraphics[width=8.5cm, height =6.0cm]{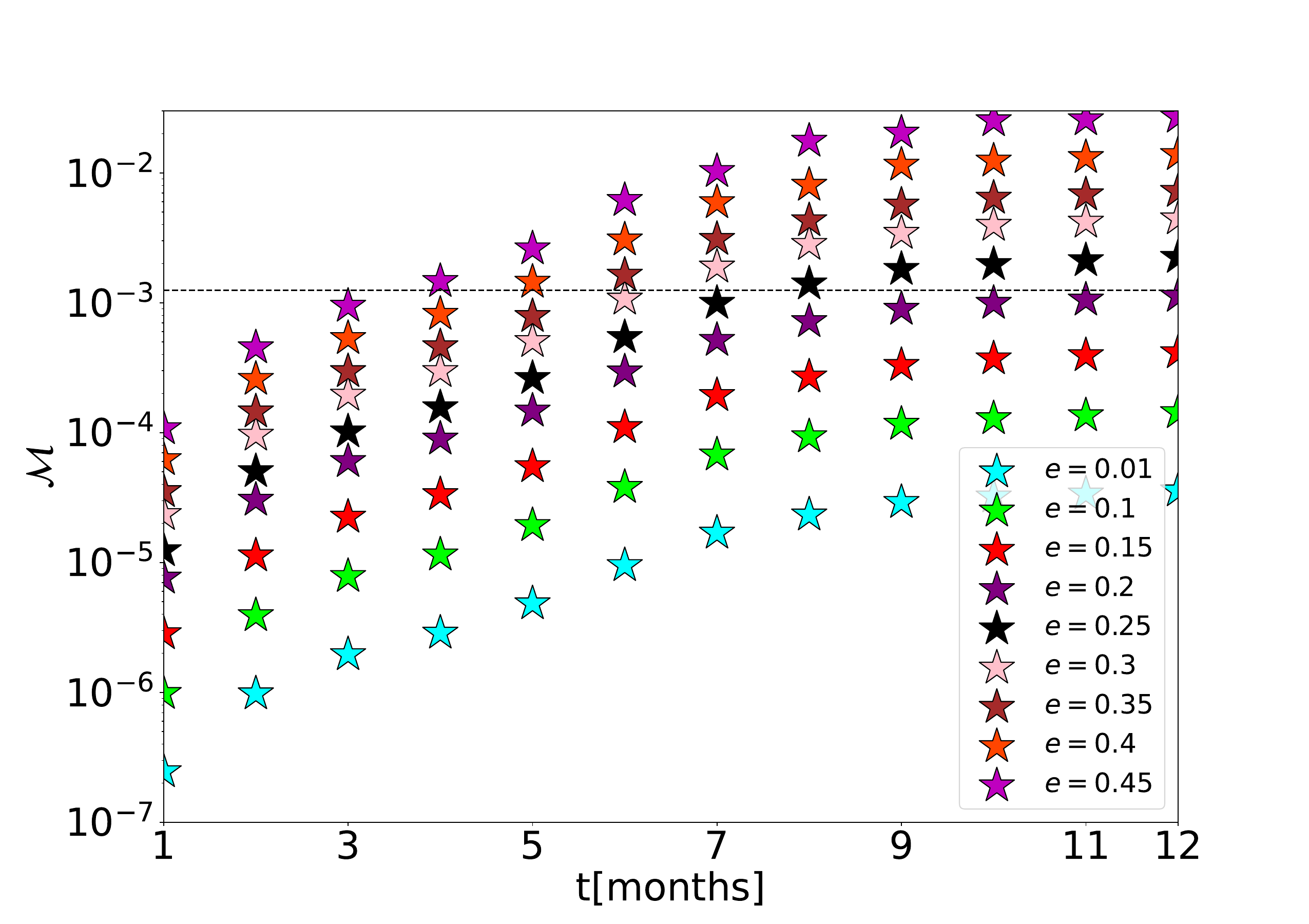}
\caption{Mismatches between EMRI waveforms from Kerr BH with and without the quantum-corrected parameter as a function of observation time are plotted, where the length of two set of waveforms are both set as one year and the spinning parameter is $a=0.1$. The left panel shows some cases of different eccentricities $e_0\in\{0.01,0.1,0.15,0.2,0.25,0.3,0.35,0.4,0.45\}$ for the  case  of parameter $L_q=2\times10^{-6}$ and semi-latus rectum $p_0=16$ , and the right panel shows examples of different deviation parameters $L_q \in\{10^{-7},5\times10^{-7},10^{-6},2\times10^{-6},3\times10^{-6},5\times10^{-6},8\times10^{-6},
10^{-5},2\times10^{-5},3\times10^{-5},5\times10^{-5}\}$ for the initial orbital eccentricity $e_0=0.2$ and semi-latus rectum $p_0=16$.
The horizontal black dashed line denotes the minimum value distinguished by LISA.}
\label{fig:mismatch}
\end{figure*}

Third, we assess the difference between EMRI waveforms of the Kerr black hole and LQGBH by computing the mismatch as a function of observation time in Fig. \ref{fig:mismatch}, where the length of two types of EMRI waveforms are set to one year. The left panel displays the mismatch  for different quantum-corrected parameter $L_q \in \{10^{-7},5\times10^{-7}, 10^{-6}, 2\times10^{-6}, 5\times10^{-6}, 10^{-5},
2\times10^{-5},5\times10^{-5},10^{-4}\}$ and a fixed orbital eccentricity $e=0.2$.
The horizontal black dashed line represents the threshold value $\mathcal{M}_c=0.00125$, above which LISA can discern the modified effect of LQGBH on the EMRI waveforms. An obvious point is that the mismatch is bigger when the parameter $L_q$ is larger; the minimum value recognized by LISA is about $2\times10^{-6}$.
The right panel of Fig. \ref{fig:mismatch}  illustrates the relationship between the mismatch  and orbital eccentricity, and one can observe that the mismatch would be slightly bigger when the geodesic orbits become more eccentric. So, the role of orbital eccentricity is also not ignored when constraining LQGBH with an EMRI signal.

To analyse the effect of the spin and mass of LQGBH  on EMRI waveforms, we can compute mismatch as a function of mass $\log_{10}(M/M_\odot)$ and spin $a$ for a fixed quantum-corrected parameter $L_q=8\times 10^{-7}$  (left panel) and $L_q=2\times 10^{-6}$ (right panel). The sub-vertical black dashed lines denote the contours of mismatches; the contour of value $1.25\times 10^{-3}$ is the threshold distinguished by LISA. One can find that mismatch is heavily subjected to the mass of the LQGBH. EMRI sources of the mass $M>10^6 M_\odot$ would not be distinguished by LISA for the case of $L_q=2\times 10^{-6}$. Also, the signal from EMRI with a mass $M\gtrsim 10^{5.7} M_\odot$ is not distinguished by LISA for the case of $L_q=8\times 10^{-7}$.
From two panels, it is found that the spin of LQGBH do not significantly impact the
mismatch.
\begin{figure*}[ht]
\includegraphics[width=8.5cm, height =6.0cm]{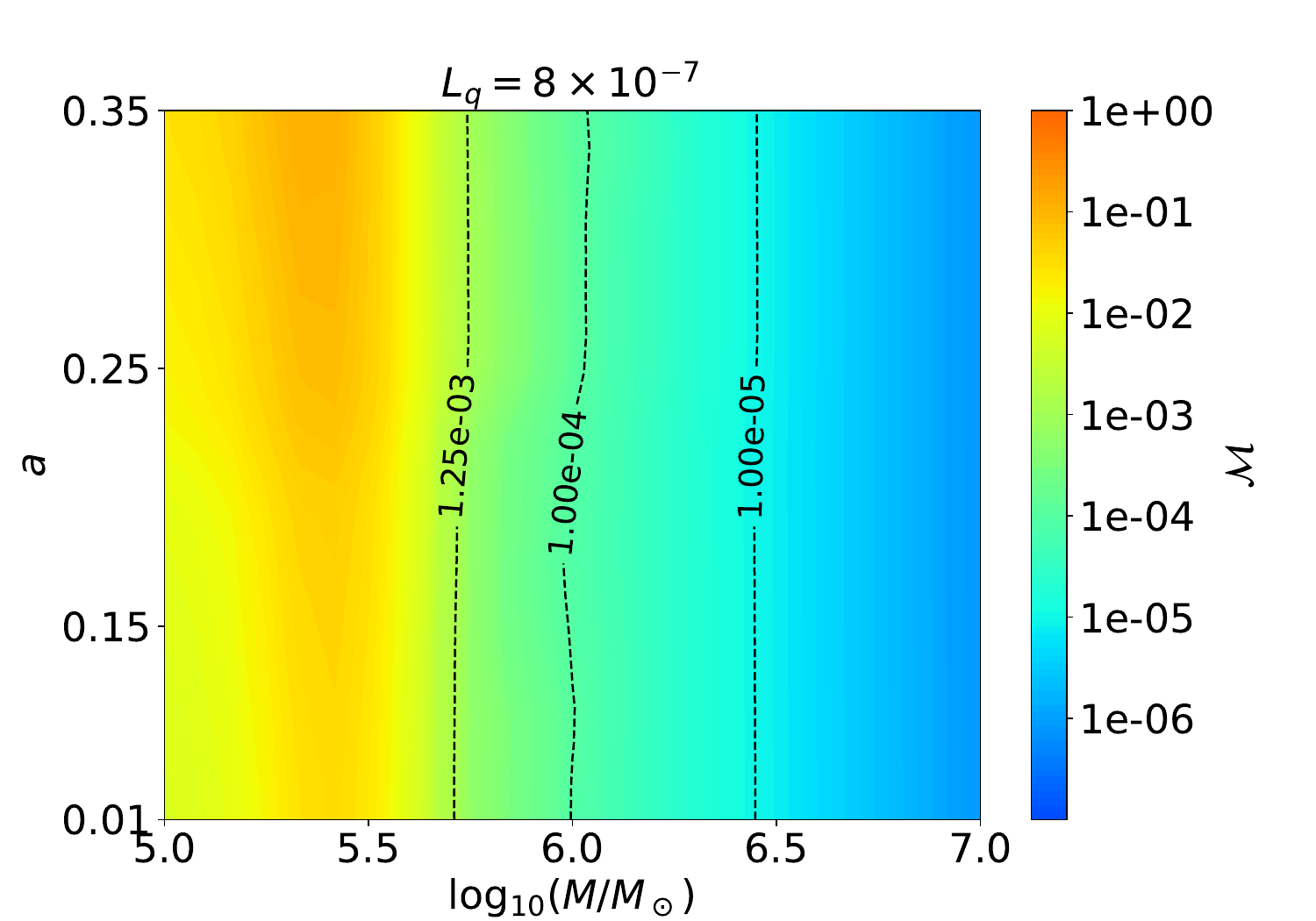}
\includegraphics[width=8.5cm, height =6.0cm]{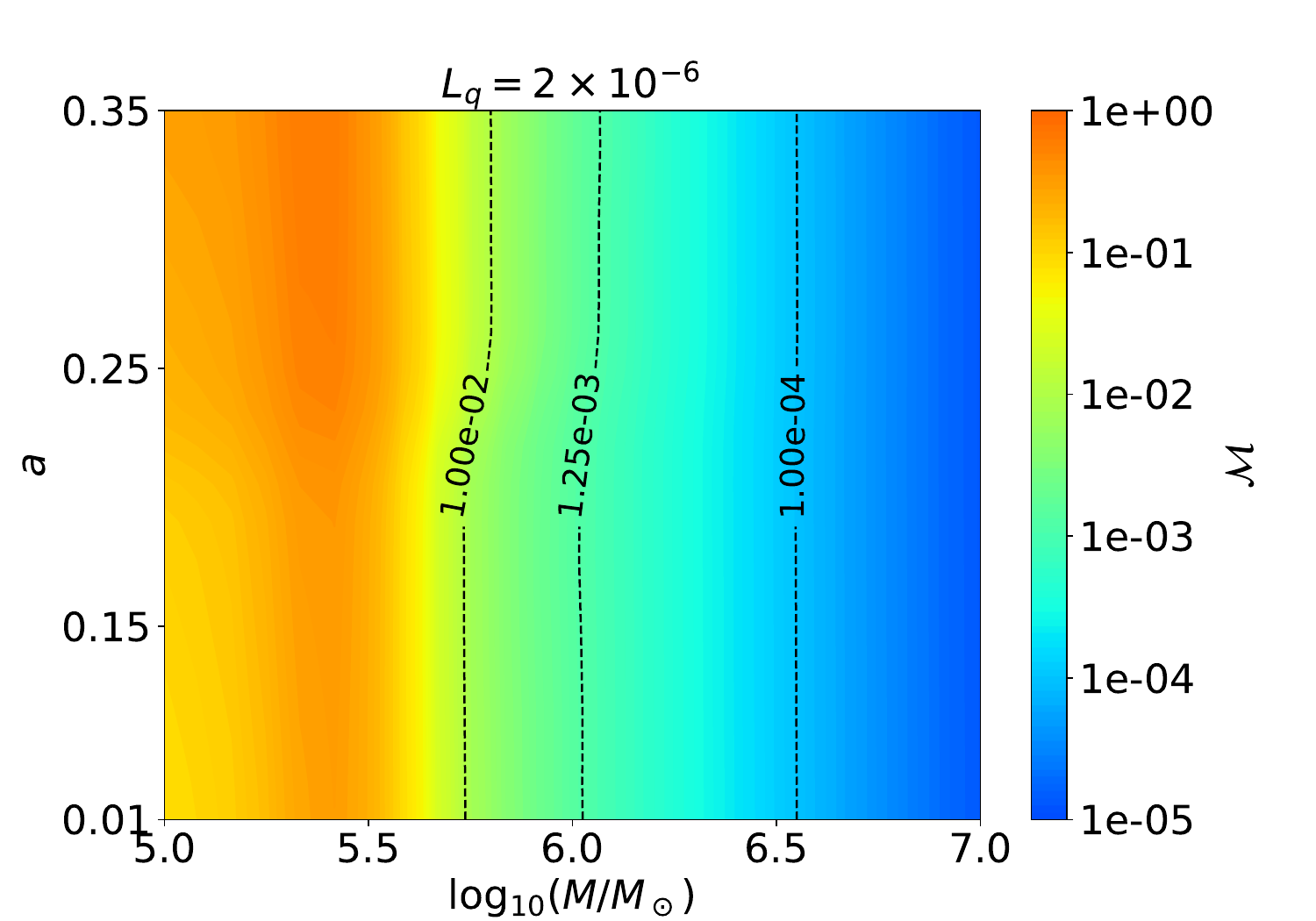}
\caption{Mismatch as a function of mass $\log_{10} (M/M_\odot)$ and  spin $a$ of LQGBH is plotted, which includes the cases of $L_q=8\times 10^{-7}$
and $L_q=2\times 10^{-6}$.
The sub-vertical black dashed lines denote to the contour value of mismatches,
among which the value $1.25\times 10^{-3}$ of contour line is just distinguished by LISA.}
\label{fig:mismatch44xcz}
\end{figure*} 

\begin{table}[h!]
\centering{\begin{tabular}{lccc}
\hline\hline
$a$          & $0.1$ & $0.1$ & $0.1$
  \\
$e_0$      & $0.1$ & $0.3$ & $0.5$
  \\
\hline\hline
$\Delta(\ln  M)$   & $1.64\times 10^{-4}$   & $7.73\times 10^{-5}$   & $1.59\times 10 ^{-5}$ 
                 \\ \hline
$\Delta (\ln m)$   & $3.54\times 10^{-4}$   & $1.23\times 10^{-4}$    & $6.92\times 10^{-5}$ 
                 \\ \hline
$\Delta(a)$       & $5.67\times 10^{-6}$   & $1.75\times 10^{-6}$     & $3.61\times 10^{-7}$
                  \\ \hline
$\Delta (e_{0})$   & $8.48\times 10^{-6}$   & $1.39\times 10^{-6}$    & $6.62\times 10^{-7}$ 
                 \\ \hline
$\Delta (p_0)$     & $1.45\times 10^{-4}$   & $2.18\times 10^{-5}$    & $8.21\times 10^{-8}$  
                  \\ \hline
$\Delta (L_q)$     & $5.46\times 10^{-6}$   & $1.82\times 10^{-6}$   & $3.11\times 10^{-7}$ 
                  \\ \hline
$\Delta(\chi_0)$   & $1.54\times 10^{-1}$  & $7.41\times 10^{-2}$    & $5.07\times 10^{-2}$ 
    \\ \hline
$\Delta(\Phi_0)$    & $2.41\times 10^{-4}$  & $2.57\times 10^{-6}$   & $6.78\times 10^{-7}$  
                 \\ \hline     
$\Delta(\Omega_S)$  & $6.09\times 10^{-3}$  & $5.41\times 10^{-3}$   & $2.41\times 10^{-4}$
                 \\ \hline
$\Delta (\Omega_K)$  & $1.25$    & $1.55$    & $8.45$
                 \\ \hline
$\Delta(\ln D)$    & $4.37\times 10^{-1}$  & $7.73\times 10^{-2}$   & $2.49\times 10^{-2}$
 \\
\hline\hline
\end{tabular}}
\caption{\protect\footnotesize
Measurement errors of EMRI parameters for the inspiral of a $30 M_\odot$
CO onto a $10^6 M_\odot$ MBH at SNR $\rho=20$.
Shown are results for various values of  the initial eccentricity $e_0$  and LQGBH
spin $a=0.1$.
The other parameters are set as follows:
$L_q=10^{-5}$,
$\Phi_0=1.0$,
$\theta_S=\pi/4$,
$\phi_S=0$,
$\chi_0=1.0$,
$\theta_K=\pi/8$,
$\phi_K=0$.
}
\label{tableII}
\end{table}

\begin{figure*}[ht]
\centering
\includegraphics[width=14.7cm, height =12.3cm]{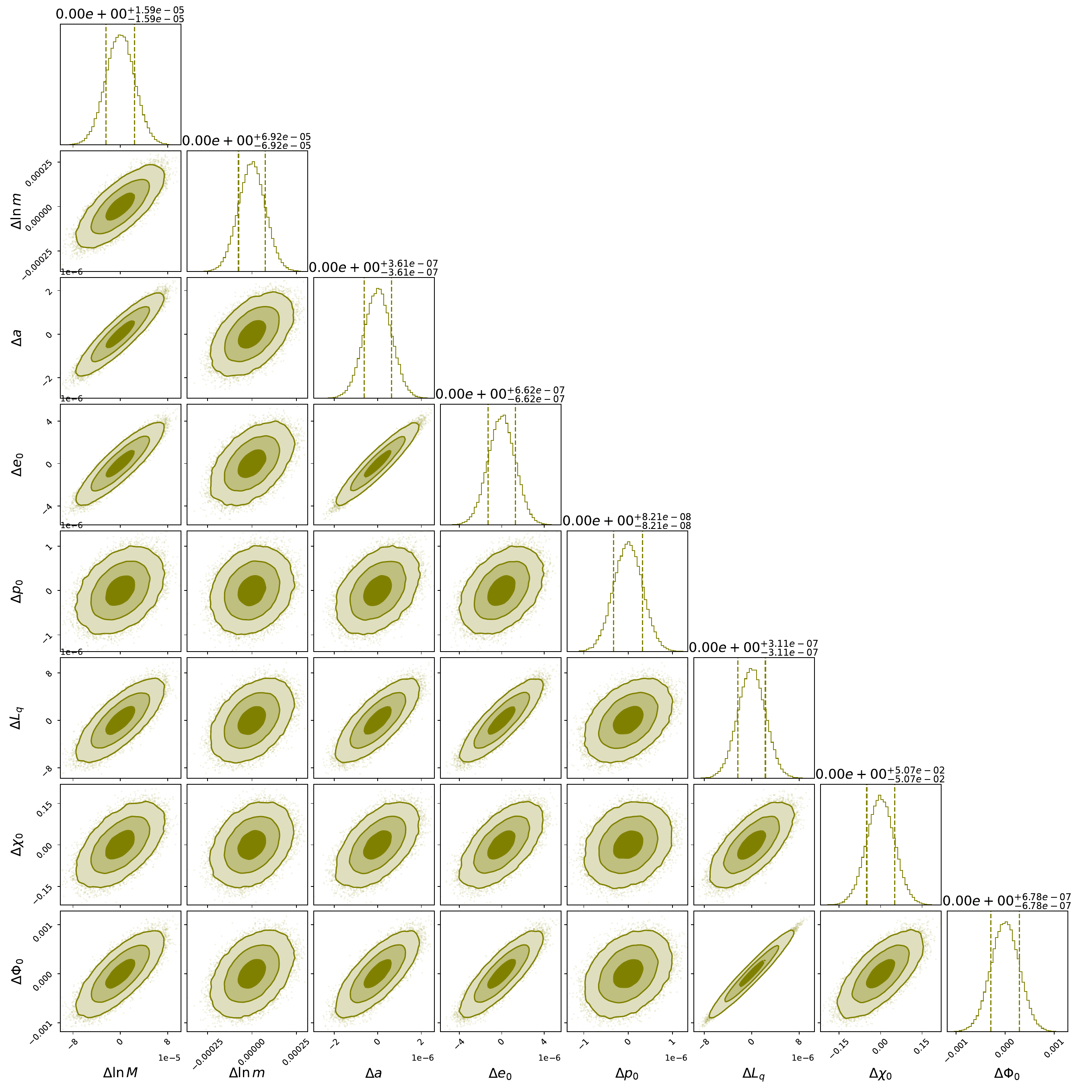}
\caption{Corner plot of the probability distribution for the mass of secondary body, mass and spin of LQGBH, initial  orbital eccentricity, quantum-corrected parameter, initial radial and azimuthal angle variables $(\ln{m}, \ln{M},  a=0.1, e_0=0.5, p_0=16, L_q, \chi_0, \Phi_0)$ is inferred from one year observation of EMRI. The vertical lines denote to $1 ~\sigma$ interval for every source parameters. Three contours show the probability confidence intervals of  $68\%$,  $95\%$ and $99\%$.}
\label{fig:conerplot}
\end{figure*}

Finally, we plan to put the constraint on the parameter $L_q$ of LQGBH by computing FIM.
Tabel \ref{tableII} lists the measurement error of EMRI source parameters for different orbital eccentricities, where the spin of LQGBH is $a=0.1$ and orbital semi-latus rectum is set to $p_0=16$. It is obvious that the measurement errors of source parameters can be slightly improved if he orbital eccentricity is bigger, the constraint on quantum-corrected parameter $L_q$ can reach a fractional error of $10^{-6}$ with the observation of LISA.
The following section analyzes the correlations among the  source parameters, for simplicity, we only show the constraint correlations with other intrinsic parameters.
Because the evolution of orbital frequencies only depends on the intrinsic parameters, as well as the GW data analysis and parameter estimation for EMRI source mainly focus on
the correlation among the intrinsic parameters \cite{Babak:2009ua,Wang:2012xh,Chua:2021aah,Katz:2021yft,Ye:2023lok,Zhang:2023vok}.
Fig. \ref{fig:conerplot} depicts the probability distribution between the
quantum-corrected parameter $L_q$ and other intrinsic parameters, which is the relatedness among parameters $(\ln{m}, \ln{M},  a, e_0, p_0, L_q, \chi_0, \Phi_0)$ using the off-diagonal elements of variance-covariance matrix.
From this Fig. \ref{fig:conerplot}, it is found that the constraint on quantum-corrected parameter $L_q$ is closely related with the parameters $(\ln{m}, \ln{M}, a, e_0, \chi_0, \Phi_0)$, and the correlation between the parameter $L_q$ and orbital semi-latus rectum $p_0$ is relatively weaker comparing with the other parameters.\vspace{-0.22cm}

\section{Discussion}\label{discussion}
GWs originate from a wide variety of sources, in which inspiralling binaries such as EMRIs are the ideal candidates for future space-based detectors. GWs generated from such sources will undoubtedly imply possible deviations from GR, giving rise to the understanding of the tests of GR and beyond by examining their strong gravity regimes. Additionally, they hold the potential to reveal various unexplored aspects of physics, such as possible quantum gravity effects, opening up fresh avenues in both GW astronomy and the foundational grasp of the GR. Hence, in this line of endeavour, the present article investigates an order of magnitude of quantum effects emerging from a loop quantum gravity-inspired black hole, also termed LQGBH, using LISA observation of EMRI waveforms.

First, starting from a rotating black hole in the LQG, we derived the analytical expressions of orbital fundamental frequencies from the geodesic equations on the equatorial plane. In the approximation of a weaker field and low velocity, we computed the analytic formulas for the rate of change of energy and angular momentum caused by the radiation reaction effects on the inspiralling object. Second, under the adiabatic approximation, we evolved the orbital parameters using energy and angular momentum losses
around the LQGBH. Third, we computed the EMRI waveforms with the quadrupole formula and then assessed the differences in EMRI signals from the Kerr black hole and LQGBH by computing the dephasing and mismatch. Finally, we conduct parameter estimation with the FIM method to constrain the LQGBH using the EMRI signal. Our results indicate that LISA would distinguish the modified effect of LQGBH with a parameter of as small as $\sim 2\times 10^{-6}$, and the constraint on LQGBH can reach a fractional measurement error of $10^{-6}$, depending on the orbital eccentricity. According to the corner plot in Fig. \ref{fig:conerplot}, the bound accuracy of the quantum-corrected parameter $\Delta L_q$ is closely related to the parameters $(\ln{m}, \ln{M}, a, e_0, \chi_0, \Phi_0)$ and is weakly correlated with the orbital semi-latus rectum $p_0$. 

Further, our results suggest a more suitable constraint on the LQG parameter ($L_q$) than the one that emerges from the black hole shadow \cite{Brahma:2020eos, Afrin:2022ztr} using the same LQG model \footnote{Refs. \cite{Afrin:2022ztr} and \cite{Brahma:2020eos} show that the upper bound on the $L_q$ from the shadow goes as $\sim 0.0821$ and, $\sim 7.7\times 10^{-5}$ respectively. Where the constraint in \cite{Brahma:2020eos} is under the solar system tests and considers the assumption that Birkhoff's theorem is still valid, which does not need to hold in LQG.}. While the photon orbits in black hole shadows are closer to the horizon, EMRIs, on the other hand, produce high-SNR, long-duration GW signals and phase evolution that accumulate over thousands of cycles, allowing for precise mapping of the spacetime geometry across a wide range of radii with eccentric orbits, especially near the LSO. Such factors, through EMRI analysis for the model under investigation, imply better constraints on non-GR parameters, as LQG effects in our context, than black hole shadow. Furthermore, we remind that, for a black hole mass of $M=10^{6}M_\odot$, the constraint $L_q= 2\times 10^{-6}$, corresponds to an LQG length scale $\sqrt{l_k}\approx 132.078$ Km. In comparison, the innermost stable circular orbit (ISCO) of a Schwarzschild black hole, located at $p_{\rm ISCO}= 6M$ translates to $p_{\rm ISCO}\approx 8.86 \times 10^6$ Km, implying that the LQG scale lies below the ISCO. The section (\ref{BHspacetime}), including \cite{Afrin:2022ztr, Brahma:2020eos}, also discusses the horizon and LSO length scales in detail. The obtained length scale implies that the constraint on the LQG parameter $L_q$ significantly deviates from the Planck length scale; we are unlikely to probe the dominant quantum gravity effects that emerge near the black hole center, i.e., comparable to the Planck length scale. Pushing $L_q<< 2\times 10^{-6}$ (equivalently, $\sqrt{l_k}<< 132.078$ Km) closer to the Planck length scale, the hunt for further minute and deeper aspects of quantum gravity effects may fall beyond LISA's detection capabilities, as per the present study. This also hints that the effective LQG spacetime used here, not strictly derived from Einstein's equations, might be better constrained using exact rotating black hole solutions in LQG models \cite{Rovelli:1997yv, Addazi:2021xuf}. Additionally, if we consider a primary mass within the typical range of LIGO/Virgo detections, say $M=36 M_\odot$ with the same $L_q= 2\times 10^{-6}$, the corresponding LQG length scale becomes $\sqrt{l_k}\approx 0.004755$ Km. This value is even smaller than the length scales associated with the masses relevant to EMRI systems stated above. Consequently, the possibility of detecting LQG effects with LIGO/Virgo is significantly diminished, as the characteristic quantum gravity length scale falls below that of EMRI systems, where detection prospects are already considered challenging. This further suggests that present ground-based GW detectors are unlikely to observe signatures of LQG corrections. On the other hand, future GW observatories, such as LISA or beyond, can be designed to detect such subtle and non-trivial effects near the Planck scale-a fundamental challenge for GW science. However, the key aspect of our EMRI-based estimate is that LQG effects, small ($\sim 2\times 10^{-6}$) but in the observable range of low-frequency detectors, can accumulate during the inspiral phase and leave detectable imprints on waveforms, offering a potential observational window with LISA.


Our current analyses pay attention only to the equatorial eccentric orbits. However, the real EMRI orbits should be inclined, so future work would consider the more general scenario with other LQG models  \cite{Rovelli:1997yv, Addazi:2021xuf}. As this is the first important step to constrain the quantum gravity parameter with GWs, implying the quantum effects, using rotating eccentric EMRI, we next aim to include the GR PN terms corresponding to the PN order of the parameter $L_q$ \cite{Moore:2016qxz}. It requires several noteworthy developments \cite{Canizares:2012is} that will be covered in later studies, leading to a complete investigation with a focus at least up to 2PN order \cite{Moore:2016qxz, Blanchet:2013haa}. Such an analysis may implicate a more stringent bound on $L_q$, and it will also serve as an extension to our present study. Further, our modeling GW signal is the condition of low frequency approximate; however, the future true EMRI detection depends on the time-delay interferometry technology \cite{Tinto:2020fcc,Tinto:2023ouy,Wang:2021owg,Tinto:2022zmf, Wang:2021jsv, Wang:2022nea, Wu:2023key}. Additionally, the constraint on LQGBH is obtained with the FIM method, so it would be necessary to infer the more rigorous results from the Bayesian Markov Chain Monte Carlo-based method \cite{Chua:2021aah,Katz:2021yft} using the time-delay interferometry signal in the forthcoming period. We aim to investigate some of these aspects in our future studies.

\section*{Acknowledgements} 
Authors thank Chao Zhang and Arpan Bhattacharyya for useful discussions. Authors also acknowledge the anonymous referee for useful comments and suggestions. T. Zi is  funded by the China Postdoctoral Science Foundation with  Grant No. 2023M731137 and the National Natural Science Foundation of China with Grants No. 12347140 and No. 12405059. The research of S. K. is funded by the National Post-Doctoral Fellowship (N-PDF: PDF/2023/000369) from the Science and Engineering Research Board (SERB-ANRF), Department of Science and Technology (DST), Government of India. The authors would also like to thank the anonymous referee for useful comments and suggestions.

\appendix

\section{Geodesic equation and constants of motion} \label{apenteu1}
Here, we present the 4-velocity of the inspiralling object in the background (\ref{met}). We keep the expressions generic in terms of angular parameter ($\theta$) and set it ($\pi/2$) at later stages for the equatorial consideration ($\mathcal{Q}=0$). We implement the Hamilton-Jacobi framework to separate out the radial and angular part of the motion \cite{Kumar:2024utz, Yagi:2023eap}
\begin{align}\label{ac1}
S = -\frac{1}{2}\mu^{2}\tau-Et+J_{z}\phi+R(r)+\Theta(\theta) \hspace{3mm} ; \hspace{3mm} -\frac{\partial S}{\partial\tau} = \frac{1}{2}g^{\mu\nu}\frac{\partial S}{\partial x^{\mu}}\frac{\partial S}{\partial x^{\nu}}.
\end{align}
Following \cite{Kumar:2024utz, AbhishekChowdhuri:2023gvu}, the separable radial and angular equations are given by
\begin{equation}
\begin{aligned}\label{j1}
\mu^{2}\rho^{4} \Big(\frac{dr}{d\tau}\Big)^{2} =& \Big((a^{2}+\omega)-a J_{z}\Big)^{2}-\Delta(\kappa+\mu^{2}\omega) \\
\mu^{2}\rho^{4} \Big(\frac{d\theta}{d\tau}\Big)^{2} =& (\kappa-\mu^{2}a^{2}\cos^{2}\theta)-\Big(aE\sin\theta-\frac{J_{z}}{\sin\theta}\Big)^{2}
\end{aligned}
\end{equation}
where $\kappa$ is the separability constant, related to a more conventional Carter constant ($\mathcal{Q}$) as $\mathcal{Q}\equiv \kappa-(J_{z}-aE)^{2}$, and ($\omega, \rho$) functions are defined in metric (\ref{met}). 
Further, the metric exhibits the two conserved quantities - energy and angular momentum - we obtain,
\begin{equation}
\begin{aligned}\label{gdscs}
\mu\frac{dt}{d\tau}  =& \frac{1}{\Delta \rho}\Big(\left(a^2+\omega\right) (a (a E-J_{z})+E \omega+a \Delta \left(J_{z}-a E \sin ^2\theta \right)\Big) \\
\mu\frac{d\phi}{d\tau} =& \frac{1}{\Delta\rho} \Big(a (a (a \text{E1}-J_{z})-E \Delta +E \omega +J_{z} \csc ^2\theta  \Delta \Big),
\end{aligned}
\end{equation}
Note that the above velocities do not consider any assumption. For computational convenience and in the spirit of analytical investigations, we assume the central supermassive Kerr-like black hole is slowly spinning . Therefore, we will take small-black hole spin approximation. With this, we can write down the following expression,
\begin{align}
\begin{split}
& \mu^{2}\Big[\Big(\frac{dr}{d\tau}\Big)^{2}+r^{2}\Big(\frac{d\theta}{d\tau}\Big)^{2}+r^{2}\sin^{2}\theta\Big(\frac{d\phi}{d\tau}\Big)^{2} \Big] = E^2-\mu ^2+\frac{4L_{q}M^{2}(3 \mu ^2 L_{q} M^2-2 Q)}{r^4} \\
& +\frac{2M(2 \mu ^2 L_{q} M^2+ Q)}{r^3}-\frac{6 \mu ^2 L_{q} M^2}{r^2}+\frac{2 \mu ^2 M}{r}
\end{split}
\end{align}
Here, we have ignored terms $\mathcal{O}(\frac{aM^{2}}{r^{4}})$ and $\mathcal{O}(\frac{aM}{r^{3}})$. It is worth mentioning that we are not approximating the parameter $L_{q}$. The only assumption is set on $a$ with the $r$ expansion about infinity as we are examining the leading-order PN analysis. Further, we can also write down the following quantity
\begin{align}
\begin{split}
& \mu^{2}\Big[\Big(\frac{dr}{dt}\Big)^{2}+r^{2}\Big(\frac{d\theta}{dt}\Big)^{2}+r^{2}\sin^{2}\theta\Big(\frac{d\phi}{dt}\Big)^{2} \Big] =  \frac{\mu ^2 \left(E^2- \mu ^2\right)}{E^2} +\frac{2\mu ^2 M \left(3 \mu ^2-2 E^2\right)}{E^2 r} \\
& +\frac{\mu ^2 M^{2} \left(12 E^2 L_{q} +4 E^2 -18 \mu ^2 L_{q} -12 \mu ^2 \right)}{E^2 r^2}
\end{split}
\end{align}
One may consider the higher-order expansion in $r$. However, it will generate sub-leading results as we are performing the leading-order PN analysis. Next, in order to separate out the rest-mass energy, we use $E=\mu+\mathcal{E}$ \cite{Misner:1973prb, Ryan:1995xi, Flanagan:2007tv, Mukherjee:2022pwd} and retain terms only up to linear order in $\mathcal{E}$. Note that we ignore terms involving $\mathcal{O}(\mathcal{E} M/r)$ as well as their higher powers. Following \cite{AbhishekChowdhuri:2023gvu},
\begin{align}\label{ener2}
\mathcal{E} =& \frac{\mu}{2}\Big[\Big(\frac{dr}{dt}\Big)^{2}+r^{2}\Big(\frac{d\theta}{dt}\Big)^{2}+r^{2}\sin^{2}\theta\Big(\frac{d\phi}{dt}\Big)^{2} \Big]-\frac{\mu M}{r}+\frac{\mu M^{2}}{r^{2}}(4+3 L_{q})
\end{align}
The higher-order terms will also contribute in (1/r) expansion. In fact, terms independent of velocities will not play a role in computing the rate of change of orbital energy and angular momentum. The reason behind this is that the loss of energy and angular momentum involves acceleration terms, which will appear from taking the time derivative of the velocities.  Further, with the linear order correction in $\mathcal{E}$, the four-velocities are given as

\begin{equation}
\begin{aligned}\label{gdscs3n}
\mu^{2}\Big(\frac{dr}{d\tau}\Big)^{2} = & \frac{12 a  J_{z} \mu  L_{q} M^2}{r^4}-\frac{4 a  J_{z} \mu  M}{r^3}-\frac{4 J_{z}^2 L_{q} M^2}{r^4}-\frac{J_{z}^2}{r^2}-\frac{6 \mu ^2 L_{q} M^2}{r^2}+\frac{2 \mu ^2 M}{r}+2 \mu  \mathcal{E} \\
\mu\frac{d\phi}{d\tau} =& \frac{J_{z}}{r^2}-\frac{6 a \mu  L_{q} M^2}{r^4}+\frac{2 a \mu  M}{r^3}-\frac{2 J_{z} L_{q} M^2}{r^4}
\end{aligned}
\end{equation}
We use these velocities to estimate the fluxes in the main text. Again, we discard subleading terms. We can further write down the angular momentum in the following manner,
\begin{align}\label{dphidt}
J_{z} = \mu r^{2}\sin^2\theta\left(1+\frac{2L_{q}}{r^{2}}+\frac{4L_{q}^{2}}{r^{4}}+\frac{8L_{q}^{3}}{r^{6}} \right)\frac{d\phi}{d\tau}-\frac{2 a E M \sin^2\theta }{r^{2}}(2 M+r).
\end{align}
The term independent of $d\phi/d\tau$ will not contribute while computing the rate of change of angular momentum since the such a calculation requires acceleration-dependent terms as mentioned previously, which will appear by taking the time derivative of the $d\phi/d\tau$ term directly. Hence, the terms, independent of $d\phi/d\tau$, will not affect the results. It is to be noted that the expressions derived, including integrals of motion, are consistent with \cite{Flanagan:2007tv, Ryan:1995xi, AbhishekChowdhuri:2023gvu, Kumar:2024utz} when $L_{q}\rightarrow 0$. 

\section{Higher-order expressions for rate change of energy and angular momentum}\label{apenf}
This presents the expressions for the average rate of change of energy and angular momentum up to 2PN order.
\begin{equation}
\begin{aligned}
<\dot{\mathcal{E}}> =& -\frac{\left(1-e^2\right)^{3/2} \left(37 e^4+292 e^2+96\right)}{15 p^5}+ \frac{24 \left(1-e^2\right)^{3/2} \left(33 e^4+104 e^2+24\right) L_{q}}{5 p^6} \\
& + \frac{a \left(1-e^2\right)^{3/2} \left(491 e^6+5694 e^4+6584 e^2+1168\right)}{30 p^{13/2}} -\frac{\left(1-e^2\right)^{3/2} L_{q}}{15 p^7} \Big[e^6 (1080 L_{q}-5789) \\
& +80 e^4 (405 L_{q}-479)+8 e^2 (7560 L_{q}-4111)+192 (54 L_{q}-19)\Big] \\
& -\frac{2 \left(1-e^2\right)^3 \left(37 e^4+292 e^2+96\right) L_q}{15 p^7} \\
<\dot{J}_{z}> =&  -\frac{4 \left(1-e^2\right)^{3/2} \left(7 e^2+8\right)}{5 p^{7/2}}+\frac{12 \left(1-e^2\right)^{3/2} \left(2 e^4+63 e^2+40\right) L_{q}}{5 p^{9/2}} \\
& +\frac{a \left(1-e^2\right)^{3/2} \left(549 e^4+1428 e^2+488\right)}{15 p^5} -\frac{2(1-e^{2})^{3/2}}{5p^{11/2}}\Big[9\left(28 e^4+297 e^2+120\right) L_{q}^2 \\
& -(480+228e^{2}+1027e^{4}+16e^{6})L_q+4(1-e^{2})^{3/2}(8+7e^{2})L_q\Big]
\end{aligned}
\end{equation}
Further, one can also write down the orbital evolution expressions in the following form
\begin{equation}
\begin{aligned}
\Big<\frac{dp}{dt}\Big> =&-\frac{8}{5p^{3}} (1-e^{2})^{3/2} \Big[(8+7e^{2})-\frac{6}{p}(24+35e^{2}+e^{4})L_{q}-\frac{a}{12p^{3/2}}(1064+1516e^{2}+475e^{4})\Big] \\
& +\frac{4 \left(1-e^2\right)^{3/2} L_q}{15 p^5} \left(48 e^6+e^4 (3101-864 L_q)+e^2 (6716-11232 L_q)+288 (7-18 L_q)\right) \\
&+\frac{16}{5p^{5}}(1-e^{2})^{2}(8+e^{2}+7e^{4})L_q \\
\Big<\frac{de}{dt}\Big> =& -\frac{e(1-e^{2})^{3/2}}{5p^{4}}\Big[(304+121e^{2})-\frac{24}{p}(93+67e^{2}+e^{4})L_q-\frac{a}{6p^{3/2}}(7032+5592e^{2}+1313e^{4}) \Big] \\
& \frac{4 e^2 (1-e^2)^2(121 e^4+183 e^2-304) L_q}{30ep^{6}}+\frac{2 (1-e^2)^{3/2} L_q}{30 e p^6} \Big[96 e^8+e^6 (11935-2808 L_q) \\
& +e^4 (43222-53136 L_q)-432 e^2 (112 L_q-57)+768\Big]
\end{aligned}
\end{equation}
\section{Stability of the Fisher matrix}\label{appB:Stability}
In this appendix, 
we use the method in Ref. \cite{Piovano:2021iwv,Zi:2023pvl}
to assess the stability of the covariance matrix after obtaining FIM.
This process of computing stability can be summarized as follows:
firstly exert a small perturbations based on the components of FIM, then see the behavior of covariance matrices. The stability can be quantitatively given by the following equation
\begin{equation}\label{fim:stab}
	\delta_{\rm stability} \equiv \mathbf{max}_{\rm ij} \left[\frac{((\Gamma+F)^{-1} - \Gamma^{-1})^{ij}}{(\Gamma^{-1})^{ij}}\right]
\end{equation}
where $F_{ij}$ is the deviation matrix, the elements is a uniform distribution $U\in[u_0, u_1]$.
To assess the stability of FIM with the EMRI signal in LQGBH spacetime, we summary the results
of stability $\delta_{\rm stability}$ in the Table \ref{FIMstability}.

\begin{table*}[!htbp]
	\caption{The stability $\delta_{\rm stability}$ of FIM with EMRI waveform from the LQGBH  with mass $M=10^{6}M_\odot$ and $L_q=10^{-5}$ is listed, considering the different spinning LQGBH.	}\label{FIMstability}
	\begin{center}
		\setlength{\tabcolsep}{5mm}
		\begin{tabular}{|c|c|c|c|}
			\hline
			\multirow{2}{*}{$U$}& \multicolumn{3}{|c|}
			{spin $a$} 
            \\
           \cline{2-4}
			& $0.1$ &$0.3$ &$0.5$ \\
            \hline
			$\in[-10^{-7},10^{-7}]$ 
           & $3.58\times10^{-2}~~$ &$4.11\times10^{-2}$ &$2.45\times10^{-2}$			
              \\		
			\hline
			$\in[-10^{-9},10^{-9}]$ & $2.43\times10^{-2}~~$ &$1.42\times10^{-2}$ &$1.56\times10^{-2}$
			             \\
			\hline	

		\end{tabular}
	\end{center}
\end{table*}

\bibliography{JN1}
\bibliographystyle{JHEP}
\end{document}